\documentstyle[12pt,aaspp4]{article}

\begin{document}
\lefthead{Montenegro, Yuan, Elmegreen}
\righthead{Curvature and Acoustic Instabilities}
\slugcomment{scheduled for ApJ, vol. 520, August 1, 1999}

\title{Curvature and Acoustic Instabilities in Rotating Fluid Disks}

\author{L.E. Montenegro}
\affil{Department of Physics, City College of New York, New York NY 10031}
\author{C. Yuan}
\affil{Institute of Astronomy \& Astrophysics, Academia Sinica, \\
P.O. Box 1-87, Nankang, Taipei, Taiwan 115, ROC}
\and
\author{B.G. Elmegreen}
\affil{IBM Research Division, T. J. Watson Research Center, \\
P.O. Box 218, Yorktown Heights, NY 10598}
\begin{abstract}
The stability of a rotating fluid disk to the formation of spiral arms
is studied in the tightwinding approximation in the linear regime.  The
dispersion relation for spirals that was derived by Bertin et al. is
shown to contain a new, {\it acoustic} instability beyond the Lindblad
resonances that depends only on pressure and rotation.  
In this regime, pressure and gravity exchange roles
as drivers and inhibitors of spiral
wave structures. Other
instabilities that are enhanced by pressure are also found in the
general dispersion relation by including higher order terms in the
small parameter $1/kr$ for wavenumber $k$ and radius $r$. We identify
two important dimensionless physical parameters:  $\epsilon=2\pi
G\sigma_0/(r\kappa^2)$, which is essentially the ratio of disk mass to
total mass (disk and halo), and $a/(\kappa r)$, which is the ratio of
epicyclic radius to disk radius ($\sigma_0$ is the mass column density,
$\kappa$ is the epicyclic frequency, and $a$ is the sound speed).  The
small term $\zeta=\left(k^2r^2+m^2\right)^{-1/2}$ is an additional
parameter that is purely geometrical for number of arms $m$. When these
terms are included in the dispersion relation, the oscillation
frequency becomes complex, leading to the growth of perturbations even
for large values of Toomre's parameter $Q$.  The growth rate is
proportional to a linear combination of terms that depend on $\epsilon$
and $a/(\kappa r)$. Instabilities that arise from $\epsilon$ are termed
{\it gravitational-curvature} instabilities because $\epsilon$ depends
on the disk mass and is largest when the radius is small, i.e., when
the orbital curvature is large.  Instabilities that arise from
$a/(\kappa r)$ are termed
{\it acoustic-curvature} instabilities, because they arise from only
the pressure terms at small $r$.

Unstable growth rates are determined for these instabilities in four
cases: a self-gravitating disk with a flat rotation curve, a
self-gravitating disk with solid body rotation, a non-self-gravitating
disk with solid body rotation, and a non-self-gravitating disk with
Keplerian rotation. The most important application appears to be as a
source of spiral structure, possibly leading to accretion in
non-self-gravitating disks, such as some galactic nuclear disks, disks
around black holes, and proto-planetary disks. All of these examples
have short orbital times so the unstable growth time can be small, even
when only terms of order $\epsilon$ contribute.

\end{abstract}


\section{Introduction}

Spiral galaxies are characterized by bright ''arms'' spiraling out from
a region near the center.  Differential rotation will shear and wind
these arms quickly if they are material features, so Lindblad (1958)
and Lin \& Shu (1964) developed a theory of density waves to overcome
this winding dilemma. Lin \& Shu (1964,1966) also obtained the
dispersion relation for these waves, which is the relation between
frequency and wavenumber.  An important discriminant in this dispersion
relation is the stability parameter $Q$ for axisymmetric disturbances
(Toomre 1964); when $Q >1$, the disk is stable against ring-like
disturbances.

Lau \& Bertin (1978) included additional terms that treated tangential
forces for fluid spiral waves in a uniform disk, finding an
additional destabilizing term they called $J$.  They used a WKB
approximation and ignored curvature terms, which scale inversely with
galactocentric radius.  Goldreich \& Lynden-Bell (1965), Zang (1976),
and Toomre (1981) also studied azimuthal forces, by considering the
temporal response of shearing wavelets.  Toomre (1981) termed the
mechanism responsible for the spectacular growth of shearing waves a
"swing amplifier".  He found that spiral waves can grow for a short
time even when $Q>1$, as long as $Q$ is not too large.

In section \ref{sect:acoustic} below, we discuss a new instability in the usual 
spiral wave equations
derived by Bertin et al. (1989; hereafter BLLT) that is relevant beyond
the Lindblad resonances, i.e., inside the ILR and outside the OLR,
even when $Q>1$.  This is a regime that BLLT did not consider.  
The new instability depends on shear and self-gravity as in the
BLLT derivation, but it also has a component in the absence of self-gravity
that arises only from pressure and rotation. 
We therefore refer to it as an {\it acoustic} instability. 

We also derive dispersion relations for fluid disks
considering the curvature terms and other terms that were ignored in
these previous studies, such as radial variations of the basic
properties of the disk. Our additional terms depend on two
dimensionless parameters, 
\begin{equation} 
\epsilon \equiv {{2 \pi G \sigma_0 }\over {r \kappa^2}}, \label{eq:epsilon} 
\end{equation} 
and $a/({\kappa} r)$, for mass column density $\sigma_0$, radius $r$,
epicyclic frequency $\kappa$, and sound speed $a$.  Typically $\epsilon
\sim 0.1$, which is small, so our new results are not important
modifications to previous studies that considered only small $Q$. However,
in regions
where $Q$ is large, the additional terms lead to residual instabilities
that can be important in some situations. 

Numerical and analytical solutions to the modified dispersion relation
are found here for typical regions in galactic and other disks.  These
include the main disks of spiral galaxies, where the rotation curves
are approximately flat (Rubin et al. 1985); the inner disks of
galaxies, where the rotation curves are approximately solid body; inner
solid-body gaseous disks that are not self-gravitating (e.g., NGC 2207;
Elmegreen et al. 1998), and non-self-gravitating Keplerian disks, as
might be appropriate for proto-planetary disks or galactic nuclear
regions surrounding black holes 
(Nakai et al. 1993).

\section{The General Dispersion Relation}
\label{sect:general}

The dynamical response of an infinitely thin fluid disk to perturbation
density waves will be studied here, considering various degrees of
approximations using algebraic expansions in terms of small
parameters.  The disk response to spiral waves is considered to be weak
enough for the linearized equations of motion to be valid.  The effects
of self-gravity, pressure, and differential rotation are included.  The
pressure is assumed to depend only on the density; in the formulation,
enthalpy is used. In the analysis, perturbation variables are assumed
to be of the form $g_{1} (r,\theta,t) = G(r) e^{i \int k(r) d r} e^{i \left(
\omega t-m\theta\right)}$, where $r$ is the radius, $\theta$ is the azimuthal
angle, $\omega$ is the frequency of oscillation if it is real, and the
growth or decay rate if it is imaginary, $m$ is the number of arms,
$k(r)$ is the radial wavenumber, and $G(r)$ is the slowly varying
amplitude.  The spiral waves have an interarm spacing that is much
shorter than the radius, that is $\zeta \equiv 1/|\hat{k}r| \ll 1$ for
total wavenumber $\hat{k} = \sqrt{k^2 + m^2/r^2}$.  This condition is
satisfied either for very short waves or for open spirals with many
arms, and it allows asymptotic solutions to the density response. The
same condition is used to express the density as a linear function of
the gravitational potential (Bertin \& Mark  1979).

The linearized equations of motion are combined with the
continuity equation to relate the perturbation
enthalpy $h_1$ to the perturbation gravitational potential $\phi_1$ 
(Goldreich \& Tremaine 1979, Lin \& Lau 1979):
\begin{equation} 
{\cal L}\left(h_1 + \phi_1\right) = -C h_1, \label{eq:operator}
\end{equation}
where ${\cal L} = d^2/{dr}^2 + A~d/dr+B$ and the coefficients are $A=
-\left(1/r\right) {d\ln{\cal A}}/{d\ln{r}}$ , $B = - m^2/r^2 +
\left(2m\Omega/{r^2 \kappa \nu}\right) {d\ln\left({\kappa^2\left(1-{\nu}^{2}
\right) / {\sigma_0 \Omega}}\right)}/{d\ln{r}}$, and ${C} =
-{\kappa^2\left(1-\nu^2\right)/a^2}$; also ${\cal A} =
{\kappa^2\left(1-\nu^2\right)/\left(\sigma_0{r}\right)}$, where  $\nu$
is the dimensionless frequency, $\nu = \left(\omega -
m\Omega\right)/\kappa$, $m$ is the number of arms in the spiral
pattern, $\kappa$ is the epicyclic frequency, $\sigma_0$ is the surface
density of the disk, $\Omega\left(r\right)$ is the angular frequency,
and $a$ is the sound speed in the disk.  The 
perturbation gravitational potential can be expressed in the form ${{\phi}_1} (r)
=\Phi(r) e^{i \int k(r) dr} $; then Poisson's equation is (Bertin \&
Mark 1979):  
\begin{equation} 
\sigma_1 = - \frac{\sigma_0}{a^2} \,f(r) \,\phi_1, \label{eq:bmpoisson}
\end{equation}
with the definition
\begin{equation}
f(r) \equiv  \frac{1}{2 \pi G r K(\alpha, m)}\,\left[ 1 + i 
\hat{A}(\alpha)\,r\, \frac{d\,\alpha}{d\,r} + \hat{B}(\alpha)\,r^2\,
\frac{d^2\,\alpha}{d\,r^2} + \hat{C}(\alpha)\,(r\,\frac{d\,\alpha}
{d\,r})^2\right], \label{eq:fterm}
\end{equation}
and the approximation
\begin{eqnarray*}
K(\alpha,m) &= &\tau\,(1 + \frac{m + 1/2}{2}\,{\tau}^2), \\
\tau &= &{({\alpha}^2 + (m+1/2)^2)}^{-1/2}, \\
\alpha &= &k\,r -i r {\Phi}^{'}/\Phi - i/2. \\
\end{eqnarray*}
This expansion for $f(r)$ is correct to third order in $\alpha$. The
terms $\hat{A}$, $\hat{B}$, and $\hat{C}$ are defined in Bertin \& Mark (1979); 
they are:  
\begin{eqnarray*}
\hat{A}(\alpha) &= &K_2 - {K_1}^2 \\
\hat{B}(\alpha) &= &{K_1}^3 +K_3 - 2 K_1 K_2 \\
\hat{C}(\alpha) &= &9{K_1}^2 K_2 - 6 K_1 K_3 + 3 K_4 - 3{K_1}^4 - 3 {K_2}^2, \\
\end{eqnarray*}
where 
\[K_{n}  = \frac{1}{n ! \,K(\alpha,m)}\frac{{\partial}^{n}\,
K(\alpha,m)}{\partial\, {\alpha}^{n}}. \]

The enthalpy, $h_1= a^2\,{\sigma_1}/{\sigma_0}$, can be expressed 
in terms of the potential $\phi_1$ using
equation (\ref{eq:bmpoisson}) to obtain
\begin{equation}
h_1 = - f(r)\,\phi_1. \label{eq:poisson}
\end{equation}
The expression for $f$, equation (\ref{eq:fterm}), can be expanded in the
small parameter $\zeta$ to get $f(r) = (\hat{k}/k_{J})\left[1 + i\,f_1\,\zeta + f_2 {\zeta}^2 +
\left(f_3+i\,f_4\right)\,{\zeta}^3 + ...\right]$. Here, $k_{J} \equiv 2
\pi G \sigma_0/a^2$ is the two-dimensional equivalent of the Jeans
wavenumber. The terms $f_i$ are real and depend on derivatives of $k$
and $\Phi$. For example, $f_1 = (k/\hat{k})\left[-1/2 -
r\,{\Phi}^{'}/\Phi - \left(1 + r\,k^{'}/k\right)\left(m^2/2
{\hat{k}}^2\, r^2\right)\right]$.  

If only the first term is kept in
the expansion of $f(r)$ and all radial gradients and the $m$-dependence
of $\hat{k}$ is dropped, the Lin \& Shu (1966) dispersion relation is
obtained:

\begin{equation} 
\left(\omega - m\Omega\right)^2 = \kappa^2-{2}\pi{G}\sigma_0|k| +k^2
a^2 .
\end{equation} 
In terms of the dimensionless frequency, $\nu = {\left(\omega -
m\Omega\right)}/{\kappa}$, dimensionless wavelength, $\eta = k_{crit}/|k| \geq
0$, where $k_{crit} = \kappa^2 /{2}\pi{G}\sigma_0$, and Toomre's stability
parameter $Q = \kappa a/{\pi{G}\sigma_0}$, the Lin-Shu relation is
\begin{equation}
\nu^2 = 1 - {1 \over {\eta}} +{{Q^2} \over {4 {\eta}^2}}.  \label{eq:linshu} 
\end{equation}

\section{Tangential forces and the stability parameter $J$}
\subsection{The Bertin-Lin-Lowe-Thurstans dispersion relation}

In the derivation of the Lin-Shu dispersion relation, which is
equation (\ref{eq:linshu}) above, terms of magnitude $m/kr$ are ignored.
Thus the dispersion relation is accurate for radial oscillations only.
When the azimuthal wavenumber $m/r$ is included, the gravitational
instability is stronger (Lau \& Bertin 1978). In the derivation of the
corresponding dispersion relation, Lau and Bertin made the assumptions
that in Poisson's equation the out of phase (i.e., imaginary) terms can
be ignored and the wavenumber $|k| \sim k_{J}/2$. Defining the total
wavelength to be $ \lambda_m = 2 \pi / \sqrt{k^2 + m^2/r^2} $,
Poisson's equation becomes
\begin{equation}
-\phi_1 = G \sigma_1 \lambda_m,
\end{equation}
and equation (\ref{eq:operator}) is 
\begin{equation}
\left({\sigma_1/\sigma_0}\right)_{in~ phase} =
{{h_1+\phi_1}\over{\kappa^2-\left(\omega-m\Omega\right)^2}} 
\left[- {{4 {\pi}^2}\over{\lambda_m^2}} - \frac{T_1}{(1-{\nu}^2)}\right],
\label{eq:lbT1} \end{equation}
where $T_1\,=\,-{(2m\Omega/\kappa\,r)}^{2}{(d \ln \Omega/d \ln r)}$. 
Note that the last term in the equation above contains $(1-\nu^2)$,
which was not present in (C15) in Lau \& Bertin (1978) because they
were considering solutions near corotation ($\nu\sim1$).  
However, $T_1/(1-\nu^2)$
can be derived from their equations (B6) and (B9), it comes from their
second term in equation (B9); in fact, Bertin et al. (1989) included it
in their dispersion relation.  Lau \& Bertin (1978) also dropped the
fifth term in (C14) when they derived (C15) because it is higher order
in $1/kr$. We do the same for equation (\ref{eq:lbT1}) because this
section is about the low order terms as well. We include all of these
terms in the higher order analysis in the rest of the paper.

The dispersion relation for spiral waves, which is analogous to
equation (\ref{eq:linshu}), is now
\begin{equation}
Q^2/4 = \hat{\eta} - \frac{(1 -{\nu}^2)}{{\hat{\eta}}^{-2} + J^2/
(1-{\nu}^2)}, \label{eq:lb}
\end{equation}
where $\hat{\eta} = k_{crit}/\hat{k}$ and $J^2 = T_1/k_{crit}^2$, as defined in
Bertin et al. (1989). We
call equation (\ref{eq:lb}) the Bertin-Lin-Lowe-Thurstan (BLLT) dispersion relation with
dimensionless frequency $\nu_{BLLT}$. It describes the response of a
differentially rotating disk to spiral perturbations.  Evidently, the
response is stronger than for axisymmetric perturbations by a factor
that depends on the parameter $J$.

Equation (\ref{eq:lb}) was
studied extensively by Lau \& Bertin (1978) and Bertin et al. (1989) 
in the limit when $\nu \approx 0$, which is near corotation. 
In this limit, equation (\ref{eq:lb}) predicts an instability when
the frequency is purely imaginary, and this occurs when
\begin{equation}
1 + \left(\frac{Q^2}{4 {\hat{\eta}}^2} - \frac{1}{\hat{\eta}}\right) \left(1 + J^2 {\hat{\eta}}^2 \right) < 0. \label{eq:lbunstable}
\end{equation}
For ring-like perturbations ($m=0$ and $J=0$), equation
(\ref{eq:lbunstable}) is satisfied when $Q<1$; that is, equation
(\ref{eq:lbunstable}) reduces to Toomre's (1964) instability condition,
$Q < 1$, for the axisymmetric case.

It is seen from equations (\ref{eq:linshu}) and (\ref{eq:lb}) that when the
imaginary terms in the equation of motion and Poisson's equation are
ignored (Hunter 1983), the dimensionless frequency is pure real or pure
imaginary according to the values of $\hat{\eta}$ and $Q$ and for small values
of $J^2$.  The exclusion of these imaginary terms is justified in the
limits $|kr| >> 1$ and $k_{crit} r >> 1$.  This latter quantity is
$\epsilon^{-1}$, defined by equation (\ref{eq:epsilon}).  If the complex
terms are included in the equation of motion and Poisson's equation,
then the frequency solutions are complex functions of $\hat{\eta}$ and $Q$.
In that case, the frequency $\nu$ contains a non-vanishing imaginary
part in all of the parameter space ($\eta, Q$). This means there is
always some instability present, consisting of an oscillation plus
growth, so $Q$ is not an absolute discriminant
of stability for small $J$ when higher order terms in $\epsilon$ are
included.  These new instabilities will be discussed in detail in
sections \ref{sect:hot} and \ref{sect:models}, 
but first we consider the low-order BLLT equation in the region beyond
the Lindblad resonances.

\subsection{A modification to the BLLT equation 
beyond the Lindblad Resonances}
\label{sect:acoustic}
In addition to the instability condition given by equation
(\ref{eq:lbunstable}), the BLLT dispersion relation (Eq. \ref{eq:lb})
predicts another instability when the frequency $\nu$ is complex and
has a real component with an absolute value larger than 1.

This is a different regime of position relative to the resonances than
considered by BLLT.  They were concerned mostly with instabilities
near corotation, where the waves are evanescent. For this reason, they
took $\nu\sim0$. In this section, we consider stability properties
inside the inner Lindblad resonance ($\nu<-1$) and outside the outer
Lindblad resonance ($\nu>1$), using the same order of approximation as
in BLLT.  These are regions that were considered to be damped and
radiative, respectively, in the BLLT model.  We show that the same
dispersion relations also allow solutions that grow as they oscillate,
i.e., with complex frequencies. 

The condition for this second instability may be obtained from the
square root part of the solution for $\nu^2$ in 
equation (\ref{eq:lb}), and is:
\begin{equation}
\left(\frac{Q^2}{4 {\hat{\eta}}^2} - \frac{1}{\hat{\eta}}\right) 
\left(\frac{Q^2}{4 {\hat{\eta}}^2} - \frac{1}{\hat{\eta}} - 
4 J^2 {\hat{\eta}}^2 \right) < 0. \nonumber
\end{equation}
This condition can be written in the form 
\begin{equation}
\frac{1}{\hat{\eta}} < \frac{Q^2}{4 {\hat{\eta}}^2} <  
\frac{1}{\hat{\eta}} + 4 J^2 {\hat{\eta}}^2, \label{eq:lbshear}
\end{equation}
which is the same as
\begin{equation}
\epsilon \hat{k} r < \frac{a^2 {\hat{k}}^2}{\kappa^2}  <  
\epsilon \hat{k} r + \frac{4 {s}^2 m^2}{{\hat{k}}^2 r^2} 
\label{eq:shearg}
\end{equation}
if we substitute $Q\epsilon/2=a/(\kappa r)$ and $\epsilon {\hat
\eta}=1/{\hat k}r$, and define $J^2/\epsilon^2\equiv {s}^2m^2$, where
${s} = 2 (-\Omega r {\Omega}^{'})^{1/2}/\kappa$ and is of order 1. 
Equation (\ref{eq:shearg}) is {\it a new condition for instability}. When
this condition is satisfied, the self-gravitating disk is unstable to
the growth of spiral waves. The right hand side of equation
(\ref{eq:shearg}) contains two terms.  The first term depends on the
self-gravity of the disk and the second depends on shear.  When gravity
is negligible, there is still instability from the second term, coming
entirely from pressure, shear, and Coriolis forces. We refer to this as
an {\it acoustic instability}; it has apparently not been considered
previously in the literature.

Figure 1 shows the unstable regions for a five-arm spiral ($m=5$) in
the $(k_{crit}/|k|,Q^2)$ plane from the BLLT dispersion relation,
equation (\ref{eq:lb}), considering a self-gravitating disk with a flat
rotation curve ($s^2=2$); this case is studied in more detail in the
next section. The growth rate is represented as a gray scale, and the
borders of the regions of instability are represented as lines,
obtained from the instability conditions.  The most unstable region is
in the bottom left corner of the figure, where the bottom line shows
the stability limit for the Lin-Shu dispersion relation ($m=0$), which
is obtained from equation (\ref{eq:linshu}). For $m$ and $J^2 \neq 0$
the border of this region of instability shifts to the line given by
the BLLT condition (Eq. \ref{eq:lbunstable}). The acoustic instability
is bracketed by the two upper lines described by equation
(\ref{eq:lbshear}). The lower line corresponds to $\nu^2=1$ . This
occurs at a Lindblad resonance
when the Doppler-shifted frequency of oscillation, $(\omega - m
\Omega)$, matches the epicyclic frequency, $\kappa$, which is where the
self-gravity of the disk is balanced by the pressure force (the Jeans
condition) according to equation (\ref{eq:lb}). The upper line
corresponds to $\nu^2 = 1 + 2 J^2 \hat{\eta}^2$.

We can investigate the instability conditions 
(\ref{eq:lbunstable}) and (\ref{eq:shearg}) further by writing 
the BLLT dispersion relation without self-gravity. 
This can be done by multiplying equation (\ref{eq:lb}) by 
$\epsilon^2$, and then substituting as above.
We then let $\epsilon \rightarrow 0$ to turn off gravity.
The BLLT dispersion relation becomes
\begin{equation}
\nu^4 - \left( 2 + a^2 {\hat{k}}^2/{\kappa^2}\right)\nu^2
+ 1 + \frac{a^2 \left({\hat{k}}^2 + {s}^2 m^2/r^2\right)}{\kappa^2} = 0. 
\label{eq:lb2}
\end{equation}
We combine the contributions to the dispersion relation from the sound
speed and the epicyclic frequency by defining an angle $\gamma =
\tan^{-1} (a\hat{k}/\kappa)$. We also define an angle 
$p = \pi+\tan^{-1} (m/kr)$ for $k<0$; this angle is between
$\pi/2$ and $\pi$, giving $\sin p>0$ and $\cos p<0$.  The standard
definition of a spiral arm pitch angle is $\pi-p$.  
For $\epsilon = 0$, equation
(\ref{eq:lbunstable}) is never satisfied, so the BLLT instability
disappears, as recognized by these authors. However, the acoustic
instability remains, with an instability criterion given by equation
(\ref{eq:shearg}) with $\epsilon = 0$; this is
\begin{equation}
\frac{a}{\kappa r} < {{2 {s} \sin p}\over{{\hat k}r}}={{2 {s} m}\over
{k^2r^2+m^2}}.
\label{eq:kepler7}
\end{equation}
Another way to write equation (\ref{eq:kepler7}) is to remove the explicit radial
dependence; then the instability condition becomes
\begin{equation}
\tan{\gamma} \equiv {{a{\hat k}}\over{\kappa}} < 2 {s} \sin{p} .\label{eq:shear}
\end{equation}
The left hand side of the inequality in equation (\ref{eq:shear}) is the ratio of the
length scale for the epicyclic oscillation to the interarm spacing.
This ratio has to be less than order unity for the instability to
develop, which means that there has to be room for epicyclic motions
within the distance that separates the spiral arms.  That is, spiral
waves will grow at all wavelengths that have enough room for epicyclic
motions at the local sound speed.

When equation (\ref{eq:shear}) is satisfied, {\it a non-self-gravitating
fluid disk with differential rotation will be unstable to spiral
perturbations inside the ILR and outside the OLR}.  
For a disk with solid body rotation, ${s} = 0$,
for a flat rotation curve, ${s} = \sqrt{2}$, and for a Keplerian
disk, ${s} = \sqrt{6}$, so condition (\ref{eq:shear}) is more easily
satisfied, and the growth of instabilities is stronger, with greater
shear. From equation (\ref{eq:lb2}), the phase velocity, $c_{ph}$, and the group
velocity, $c_{g}$, of the acoustic waves in the radial
direction can be obtained. Define
$z =  (2 {s} \sin{p}/ \tan{\gamma})^2$, and $w_{\pm} = 
(1 \pm \sqrt{1 - z})/2$; then 
\begin{eqnarray*} 
c_{ph} &= &m \Omega/k \pm \frac{\kappa}{k} \sqrt{1 + w_{\pm} \tan^{2} \gamma },\\ 
c_g &= &\pm a \frac{ w_{\pm} \cos{p} \tan{\gamma}} {\sqrt{1 - z} 
\sqrt{1 + w_{\pm} \tan^2 \gamma}}. \\ 
\end{eqnarray*} 
In the unstable regime, $z>1$, which implies that $c_{ph}$, and $c_{g}$
are complex; only the real parts should be taken for the physical phase
and group velocities.  Note that for trailing waves, which are the only
waves considered here, $\cos p<0$.  This instability, along with
additional instabilities resulting from higher order terms, will be studied
further in section \ref{sect:models} 
for the cases with flat rotation curves and Kepler
rotation.

\subsection{Physical insights to the BLLT extension beyond the
Lindblad resonances}
\label{sect:blltextend}

The acoustic instability determined by the condition (\ref{eq:shear})
has a different physical origin than the higher-order instability
discussed in the next sections.  For example, the higher-order
instability works with or without shear, but the BLLT extension beyond
the Lindblad resonances requires shear ($s\ne0$ in equation
\ref{eq:shear}).  We show in section \ref{sect:bessel} that the
physical origin of the higher order instability is a geometric
growth of incoming wavetrains near the nucleus of a galaxy.  We do
not actually think of this higher-order growth as an instability because
it is limited in time to the propagation time over the radius.
This is unlike the acoustic instability discussed in the previous
section, which is a true instability. The acoustic instability is very
similar to the gravity-driven instability of BLLT near corotation,
i.e., between the Lindblad resonances, but it is pressure-driven
instead, and beyond the Lindblad resonances. We explain here in
physical terms how it works.

In normal galactic spirals between the Lindblad resonances, and in bars
between corotation and the ILR, the spiral or bar perturbation grows
with time because more and more stellar (or fluid particle) 
orbits lock into phase with the
perturbation, and because each new aligned orbit reinforces the
perturbation, causing greater and greater forcing.  This works for two
reasons: (1) In this radial range, an unperturbed epicycle precesses
slower than the pattern speed, i.e., the precession speed,
$\Omega-\kappa/m$, is less than the pattern speed, $\Omega_p$.  (2) The
{\it inward} forcing from the perturbation, gravity in this case, is
greatest near the apocenter of the epicyclic orbit.  For a spiral arm,
this apocenter occurs just outside the potential minimum of the arm,
and is directed inward because of the arm gravity.  For a bar, the
apocenter is on the bar major axis, and is directed inward because of
the gravity of the bar.

Reason (1) implies that in the absence of forcing, a fluid element with its
apocenter at the crest of one arm will come in and go out again to
the next apocenter {\it before} it reaches the next arm.  This is
because the precession rate is slow and the apocenter of the epicycle
twists around in angle more slowly than the spiral pattern.  
(In other words, the Coriolis force (in $\kappa$) is too large, so the angular 
velocity perturbation causes too large a radial velocity
perturbation and the radial oscillation period is short.)
However, with gravity, 
the excess inward forcing at the apocenter in the first arm crest gives
the orbit an extra kick in the radial direction, and this flings the
fluid element all the way around to the next arm before it has its next
apocenter. Moreover, this kick occurs during the part of the 
orbit when the fluid is most susceptible to gaining momentum, i.e., when it
is moving most slowly and spending the most time (at apocenter).  Thus
the forced orbit aligns with the perturbation, always having its
apocenter in the arm crest.  The same occurs for a bar: the presence of
an excess inward bar forcing on the major axis of the bar flings the
fluid elements around so they have their next apocenter at the other major axis,
rather than too early.  Thus we see how the gravitational force from
spirals and bars causes the epicyclic motions of individual fluid elements to
align with the perturbation and strengthen it.

Inside the inner Lindblad resonance, the precession speed of an
unforced stellar orbit is {\it greater} than the pattern speed, i.e.,
$\Omega-\kappa/m>\Omega_p$, so normal spiral or bar gravity kicks the
stellar orbits the wrong way.  That is, the gravity forcing makes an
epicycle that already has its next apocenter come too late, meet the
next arm even later.  For the case of the bar, this leads to a
perpendicular alignment of the orbits, so the point of maximum inward
forcing, on the bar axis, is at the place in the epicycle, its
pericenter, where the fluid element is least susceptible to acquire excess
momentum, i.e., where it is moving most quickly.
A previous description of this process was given in Elmegreen (1997). 

Now consider the influence of pressure on these waves.  The pressure
forcing in a spiral is out of phase from the gravity forcing. When
the gravity forcing is a maximum {\it inward}, just outside the spiral
potential minimum, the pressure forcing is a maximum in the {\it
outward} direction, because of the pressure gradient from the
compression in the arm crest.  Thus pressure is a stabilizing influence
on normal spirals and bars between the Lindblad resonances, as is well
known.  Pressure forcing has to be less than gravity forcing for the
spiral to grow.  This is the usual condition for the dispersion
relation, which equates the wave oscillation frequency to positive (and
therefore stabilizing) contributions from acoustic and epicyclic
oscillations, plus a negative (and therefore destabilizing) contribution
from self-gravity.

Inside the ILR and outside the OLR, the role of pressure and gravity change.
Whereas self-gravity opposes the alignment of epicycles beyond the
Lindblad resonances, as discussed above, pressure is in the right
phase to support this alignment of epicycles. The maximum
outward force from pressure is near the epicycle apocenter both inside
and outside the ILR, and the existence of this outward force slows down the
fluid at its apocenter in both cases too. But inside the ILR and outside
the OLR, this slow down causes the next apocenter to occur in the next
arm, rather than after the next arm, which would be the case without
the pressure forcing.

The acoustic instability beyond the Lindblad resonances
is therefore due to a reversal in the role of gravity and pressure as
driving agents for spiral density waves on either side of the Lindblad
resonances.  Between the ILR and OLR, gravity changes the orbits in such
a way that they reinforce an initial perturbation, while pressure
opposes this change. Beyond the Lindblad resonances, pressure
changes the orbits to reinforce the initial perturbation, while gravity
opposes. When gravity is weak beyond the Lindblad resonances, pressure
alone is left to drive spiral instabilities.

The sensitivity of the instability condition (\ref{eq:shear}) to shear
($s$) and pitch angle ($\pi-p$), which is the same
as the requirement that $\nu^2<1+2J^2{\hat \eta}^2$,
makes sense for such pressure driven
spirals.  When the pitch angle is large, the maximum inward pressure
force occurs closest to the minor axis of the epicycle, and the maximum
outward pressure force occurs closest to the major axis. This situation
leads to the maximum possible forcing from the pressure gradients. The
shear is important because this is what causes the epicycles to precess
forward or backward relative to the pattern. Without shear, the
precession speed is zero, and no amount of pressure forcing can enhance
the spiral alignment of orbits.

\section{Higher Order Terms in the General Dispersion Relation}
\label{sect:hot}

For the general case with self-gravity, it is possible to solve for the
complex frequency if we know the basic state of the disk.  If the
rotation curve, the density distribution, and the sound speed
distribution in the disk are known, then the dispersion relation in the
tightwinding approximation can be obtained to second order in
$\epsilon$.

The dispersion relation for $\nu$ is obtained by turning 
equation (\ref{eq:operator}) into an 
algebraic expression. This is done by using the definition of the enthalpy
and equation (\ref{eq:bmpoisson})
to express the enthalpy as a function of the potential and then using
the asymptotic form of the potential, $\phi_1 = {\Phi} e^{i\,\int 
{k(r)\, dr }} $. We will consider only trailing spirals ($k < 0$).  
Note that ${\nu}^{'} = -(m{\Omega}^{'}/\kappa + \nu \kappa^{'} /\kappa)$
for radial derivatives denoted by primes. 
Equation (\ref{eq:operator}) can be written in the form:
\begin{equation}
\left[\frac{r^2\,d^2}{{dr}^2} + \left(A\,r\right) \frac{r\,d}{dr}+B\,r^2\right]
\left(h_{1}+ \phi_{1}\right) = {\delta}^{-1}\,(1 - {\nu}^2) h_1, \label{eq:op1}
\end{equation}
where $\delta = a^2/({\kappa}^2\,r^2)$.
Multiply equation (\ref{eq:op1}) by $\delta/{h_1}$ and define 
$D_0 = \delta \,(1-\frac{1}{f})$, 
$D_1 = \delta \,\frac{r}{h_1}\frac{d}{dr}(\phi_1+h_1)$, and 
$D_2 = \delta \,\frac{1}{h_1}(\frac{r^2\,d^2}{dr^2} - m^2) (\phi_1 + h_1)$.
Then 
\begin{equation}
D_2 + (A\,r)\,D_1 + (B\,r^2 + m^2)\,D_0 + {\nu}^2-1 = 0. \label{eq:disp}
\end{equation}
The terms $D_i$ are 
\begin{eqnarray*}
D_0 &= & \delta\left(1 - {{1} \over {f}}\right),\\ 
D_1 &= & \delta\,\left[\left(i\,k\,r + {{r {\Phi}^{'}}\over{\Phi}}\right)\left(1 - {{1}\over{f}}\right) + 
 {{r f^{'}}\over{f}}\right] \\ 
D_{2} &= & \delta\,\left(
\left[-{\hat{k}}^2 r^2 + i k r\,
\left({{2\,r {\Phi}^{'}}\over{\Phi}} + {{r\,k^{'}}\over{k}}\right) 
+ {{r^2 {\Phi}^{''}}\over{\Phi}}\right] 
\left(1 - {{1}\over{f}}\right) 
+ {{2 r f^{'}}\over{f}} 
\left(i k r + {{r {\Phi}^{'}}\over{\Phi}}\right) 
+ {{r^2 f^{''}}\over{f}}\right), \\
\end{eqnarray*}
These terms are used to find numerically the roots of the dispersion relation.
They can be expanded in the small parameter $1/{\hat k} r$ by using
the Bertin \& Mark expression of Poisson's equation (Eqs. \ref{eq:fterm}
and \ref{eq:poisson}).
Their expansion is correct to third order in $1/{\hat k} r$, so our
dispersion relation is limited to third order in this quantity as well.
In terms of $Q^2$ and $\hat{\eta}$, and to lowest order in $\epsilon$, $D_i$
become:
\begin{eqnarray*}
D_0 &= & {\epsilon}^{2} \left({{Q^2}\over{4}}- \hat{\eta}\right) + i\,{\epsilon}^3 
{\hat{\eta}}^2\,f_1 + ... 
    =  \epsilon^2 d_{0\,2} + i\,\epsilon^3 d_{0\,3} + ... \\
D_1 &= & i\, \epsilon \cos{p}\,({{Q^2}\over{4 \hat{\eta}}} - 1) +
{\epsilon}^2\,\left[{{Q^2}\over{4}}\,\left({{r\,{\Phi}^{'}}\over{\Phi}} + 
{{r\,{\hat{k}}^{'}}\over{\hat{k}}} - {{r\,{k_{J}}^{'}}\over{k_{J}}}\right) + 
\hat{\eta}\,\left(f_1 \cos{p} - {{r\,{\Phi}^{'}}\over{\Phi}}\right)\right] + ... \\
    &= & i\,\epsilon d_{1\,1} + \epsilon^2 d_{1\,2} + ... \\
D_2 &= & {{1}\over{\hat{\eta}}} - {{Q^2}\over{4\,{\hat{\eta}}^{2}}} + 
i\,\epsilon\, \left[ \cos{p} \left( {{Q^2}\over{4\,\hat{\eta}}} - 1
\right) \left( {{2\,r\,{\Phi}^{'}}\over{\Phi}} + {{r\,k^{'}}\over{k}} \right) 
- f_1 +\cos{p}\,{{Q^2}\over{2\,\hat{\eta}}}
\left( {{r\,{\hat{k}}^{'}}\over{\hat{k}}} - {{r\,{k_J}^{'}}\over{k_J}} \right)
\right] \\
    &  & + {\epsilon}^2\, \left( {{Q^2}\over{4}}{{r\,{\Phi}^{''}}\over{\Phi}} 
- {\hat{\eta}}\, \left[ {{r\,{\Phi}^{''}}\over{\Phi}} + \cos{p}\,f_1\,
\left( {{2\,r {\Phi}^{'}}\over{\Phi}} + {{r\,k^{'}}\over{k}} \right) 
+{f_1}^2 + {f_2} \right] \right) \\
    &= & d_{2\,0} + i\,\epsilon d_{2\,1} + \epsilon^2 d_{2\,2} + ... \\
\end{eqnarray*}
These equations define the terms $d_{i\,j}$; note that alternate terms
are imaginary as is typical for WKB approximation methods.  Also note that $r{\hat{k}}^{'}/\hat{k} = \cos^2 {p}\, r
k^{'}/k - \sin^2 {p}$, and that $r {k_J}^{'}/k_J = r
{\sigma_0}^{'}/\sigma_0 - 2 r a^{'}/a$.  
We take $k$ to be constant and real.
The terms ($A\,r$) and ($B\,r^2 + m^2$) contain contributions of order
unity divided by $\nu$ and $({\nu}^2 - 1)$. To get a polynomial
expression for $\nu$, we calculate the expressions
\begin{eqnarray*}
\nu\,({\nu}^2 -1)\,A\,r &= & a_1\,\nu + a_2\,{\nu}^2 + a_3\,{\nu}^3 \\
\nu\,({\nu}^2 -1)\,(B\,r^2 + m^2) &= & b_0 + b_1\,\nu + b_2\,{\nu}^2 \\
\end{eqnarray*}
with
\begin{eqnarray*}
a_1 &= & {{2\,r\,{\kappa}^{'}}\over{\kappa}} - 1 - {{r\,{\sigma_0}^{'}}\over{\sigma_0}} \\
a_2 &= & {{2\,m\,\Omega}\over{\kappa}}{{r\,{\Omega}^{'}}\over{\Omega}} \\
a_3 &= & 1 + {{r\,{\sigma_0}^{'}}\over{\sigma_0}} \\
b_0 &= & {{2\,m\,\Omega}\over{\kappa}}\,({{r\,{\sigma_0}^{'}}\over{\sigma_0}} + 
{{r\,{\Omega}^{'}}\over{\Omega}} - {{2\,r\,{\kappa}^{'}}\over{\kappa}}) \\
b_1 &= & - \left({{2\,m\,\Omega}\over{\kappa}}\right)^2 {{r\,{\Omega}^{'}}\over{\Omega}} 
\equiv J^2/\epsilon^2  = {s}^2 m^2 \\ 
b_2 &= & - {{2\,m\,\Omega}\over{\kappa}}\,({{r\,{\sigma_0}^{'}}\over{\sigma_0}}
 + {{r\,{\Omega}^{'}}\over{\Omega}}). \\
\end{eqnarray*}

Equation (\ref{eq:disp}) is now multiplied by $\nu(\nu^2-1)$ to obtain a general 
dispersion relation for fluid disks: 
\begin{equation}
\nu^5 + c_3 \nu^3 + c_2 \nu^2 + c_1 \nu + c_0 = 0,  \label{eq:fifth} \\
\end{equation}
where
\begin{eqnarray*}
c_3 &= & -2 + d_{2\,0} + i\,\epsilon\,(d_{2\,1} + a_3 d_{1\,1}) + \epsilon^2\,
(d_{2\,2} + a_3 d_{1\,2})  + ... \\
    &= & c_{3\,0} + i\,\epsilon c_{3\,1} + \epsilon^2 c_{3\,2} + ... , \\
c_2 &= & i\,\epsilon a_2 d_{1\,1} + \epsilon^2 (a_2 d_{1\,2} + b_2 d_{0\,2}) + ... \\
    &= & i\,\epsilon c_{2\,1} + \epsilon^2 c_{2\,2} + ..., \\
c_1 &= & 1 - d_{2\,0} + d_{0\,2} J^2 + i\,\epsilon (-d_{2\,1} + a_1 d_{1\,1} + 
J^2 d_{0\,3}) + ... \\
    &= & c_{1\,0} + i\,\epsilon c_{1\,1} + \epsilon^2 c_{1\,2} + ... , \\
c_0 &= & \epsilon^2 b_0 d_{0\,2} + i\,\epsilon^3 b_0 d_{0\,3} + ... \\
    &= & \epsilon^2 c_{0\,2} + i\, \epsilon^3 c_{0\,3} + ... . \\
\end{eqnarray*}
This dispersion relation includes terms that have been neglected in previous studies.
The effect of the higher order terms can be followed by the dependence of the 
coefficients $c_i$ on the small parameter $\epsilon$. In the limit of $\epsilon 
\rightarrow 0$, but with a finite $\hat{\eta}$ and finite $Q^2/{\hat{\eta}}^2$, the general 
dispersion relation (Eq. \ref{eq:fifth}) becomes the BLLT dispersion relation 
(Eq. \ref{eq:lb}). 

We investigate the effects of the higher order terms by expressing $\nu$ as an 
expansion in the parameter $\epsilon$, that is, $\nu = \nu_0 + \epsilon \nu_1 + 
\epsilon^2 \nu_2 + ...$, and by solving for the roots of equation (\ref{eq:disp}). 
Substituting the expansion for $\nu$ into equation (\ref{eq:disp}) and setting 
coefficients of equal powers of $\epsilon$ to zero, we obtain expressions for the 
expansion terms $\nu_i$.
The zeroth-order root, $\nu_0$, satisfies the equation
\begin{equation}
{\nu}_0\, \left[{\nu_0}^4 - {\nu_0}^2\,\left(2 - d_{2\,0}\right) + 
1 -d_{2\,0}\,\left(1 +
J^2\, {\hat \eta}^2\right)\right] = 0. \label{eq:zeroth}
\end{equation}
The expression in the squared brackets of equation (\ref{eq:zeroth}) is the 
BLLT dispersion relation as discussed above. The
other solution ($\nu_0\,=\,0$) has no terms of order $\epsilon$; i.e., it is
of the form $\nu\,=\,{\nu}_2\,{\epsilon}^2
+ {\nu_3}\,{\epsilon}^3 + ...$.

The first-order term that corresponds to the nonzero solution $\nu_0$ is
\begin{equation}
{\nu}_1 = -i\,\frac{\nu_0\left(c_{1\,1} + c_{2\,1}\,\nu_0 + c_{3\,1}\,
{\nu_0}^2\right)}{c_{1\,0} 
+ 3 c_{3\,0}\,{\nu_0}^2 + 5 {\nu_0}^4}; 
\nonumber
\end{equation}
for real $\nu_0$ (i.e., stability in the BLLT equation), this $\nu_1$ is 
purely imaginary; for imaginary $\nu_0$, it is complex.

The coefficients, $c_{i\,2}$, for the next term in the expansion,
$\nu_2$, are pure real and, if $\nu_0$ is real, then this term is real
also, and a factor of $\epsilon^2$ smaller.  When $\nu_0$ is real, the
growth rate to first order in $\epsilon$ is attributed to 
$\nu_1$. The next contribution to the growth rate will be from
$\nu_3$, which is of order $\epsilon^2$ smaller.

In summary, we have found in this analysis a general dispersion
relation that includes the effects of radial variations in the basic
parameters of the disk and is accurate to higher order
in the small parameter $\epsilon=\left(k_{crit}r\right)^{-1}$.
Furthermore, the effects included in this
analysis change significantly the criterion for stability of the disk
as shown explicitly by the models in the next section.

\section{Instability models including the high order terms}
\label{sect:models}

Several models will be studied to illustrate the effects of the higher
order terms in the dispersion relation and to investigate how different
assumptions affect the stability of the disk.  Four models will be
considered: a self-gravitating disk with a flat rotation curve, a
self-gravitating disk with solid body rotation, a non-self-gravitating
disk with solid body rotation, and a non-self-gravitating disk with
Keplerian rotation.  The amplitude of the wave is assumed to be slowly
varying so $r\,{\Phi}^{'}/\Phi \ll 1$.  This gives an arm/interarm
contrast that increases with radius beyond one scale length, in
agreement with observations (Schweitzer 1980; Elmegreen \& Elmegreen
1984).

All disks considered here are assumed to have an exponential mass
column density profile with a scale length $r_{d}$ and a constant
sound speed, $a$.  Then $r\,{\sigma_0}^{'}/{\sigma_0} = -r/r_{d}$
and $r\,a^{'}/a=0$.

We are considering solutions to the dispersion relation obtained
from a local analysis where there are gradients in the physical
quantities of the equilibrium disk. The local analysis is relevant when
the growth time for the perturbations is shorter than the time needed
for the disturbances to travel to the boundaries (e.g., see
Lin \& Shu 1964; Toomre 1981). That is usually
$10^9$ years to the outer boundary and $10^7$ years to the center for a
circumnuclear disk, but in this case the center boundary usually serves
as a sink, as waves are shocked and energy is dissipated. Therefore we
are justified in using a local analysis in nuclear disks. For main
galaxy disks the growth time of spiral waves is also typically less
than the propagation time.
Bertin et al. (1989) considered a non-local analysis, including the
effects of gradients and boundary conditions. This leads to the
standard modal theory of spiral structure.

Gradients of disk properties, as well as curvature, can lead to spatial
variations in the amplitude of spiral waves, including singularities.
The curvature effects are considered in more detail in section 
\ref{sect:bessel}.

General dispersion relations like these 
can be solved by assuming $k$ real
and $\omega$ complex, or $k$ complex and $\omega$ real.
In the
remainder of this section, we consider $k$ real and constant and look
for solutions with imaginary $\omega$. The result will be sinusoidal
waves that grow exponentially with time, as in the usual stability
analyses.

A third method of analysis is to consider the initial value problem of
time-dependent growth with shearing sinusoidal perturbations, as in
Goldreich \& Lynden-Bell (1965) and Toomre (1981). When gravity is
important, this leads to the swing amplifier theory.

In the following subsections we will investigate analytically and
numerically the dispersion relation for disks with different rotational
properties. The relevant dispersion relation is Eq.  (\ref{eq:disp}).
The same dispersion relation with explicit expansion in terms of the
small parameter $\epsilon$ is Eq. (\ref{eq:fifth}) for self-gravitating
disks. Another dispersion relation is derived for non-self-gravitating
disks in the indicated subsections.

\subsection{Exponential self-gravitating disk with constant rotation velocity}
\label{sect:sgconstant}

We first find the roots of Eq. (\ref{eq:disp}) at two scale lengths for
an exponential disk with a constant rotation speed.  In this case $r\,
{\Omega}^{'}/{\Omega} = r\,{\kappa}^{'}/\kappa = -1$, and ${\Omega/{\kappa}} 
\,=\,1/\sqrt{2}$.  The value of $\epsilon=1/(k_{crit}r)$ depends on the
ratio of the disk to total mass (disk and halo) in the spiral region. A
value of $\epsilon \sim 0.11$ corresponds to the Solar radius in the
Galaxy, using the rotation curve model in Schmidt (1983) and a disk
mass surface density of $48$ M$_\odot$ pc$^{-2}$ (Kuijken \& Gilmore
1989, 1991). We use a value of $\epsilon = 0.1$.

There are five roots of the dispersion relation. The root that
corresponds to the greatest growth is always plotted in the figures
here; this is the root with most negative imaginary component.

Figure 2 shows the components of the normalized frequencies $\nu$ in
the ($k_{crit}/|k| \equiv \eta$, $Q^2$) plane for two values of the
azimuthal wavenumber, $m = $2, and 5, obtained numerically using the
full dispersion relation, equation (\ref{eq:disp}) with coefficients up
to third order in the small parameter $\zeta \equiv \epsilon {\hat
\eta}$. To be clear, we write $k_{crit}/|k|$ instead of $\eta$ in the
figures. The top figures show the negative of the imaginary component
of the frequency, i.e., the growth rate normalized to the epicyclic
frequency $\kappa$, with contour values $2^{i/4}$ for $i =$ -20 to 10.
The bottom figures show the absolute values of the corresponding real
frequencies with the same contours.  The left figures correspond to
$m=$ 2 and the right correspond to $m=$ 5. The values of the real and
imaginary components are tabulated for some values of $Q^2$ and
$k_{crit}/|k|$ in table 1; this will facilitate the interpretation of
the contours.

The thick lines in the top plots of figure 2 indicate the loci of
points where the normalized frequency, $\nu_{BLLT}$, equals 0
(corotation, lower line), $\pm1$ (inner and outer Lindblad resonances,
middle line) and $\pm \sqrt{1 + 2 J^2 {\hat{\eta}}^2}$ (upper line) in
the BLLT dispersion relation, equation (\ref{eq:lb}). The BLLT
instability condition, equation (\ref{eq:lbunstable}), is satisfied
below the lower thick line.  The new acoustic instability condition,
equation (\ref{eq:shearg}), is satisfied between the middle and the
upper thick lines.

The figure and table show that the growth rate decreases but remains
finite for $k_{crit}/|k| \rightarrow 0$, and that at $k_{crit}/|k| =
0$, it increases with increasing $Q$.  At intermediate values of
$k_{crit}/|k|$, say 0.5, the growth rate is largest for $Q<1$ and
decreases to a minimum at $Q^2 \approx 2$, but again increases for
increasing $Q^2$. The growth rate decreases for increasing
$k_{crit}/|k|$ beyond 0.5 for constant values of $Q^2$. This pattern is
observed for both $m$ values. A significant difference between the
figures for $m=2$ and $m=5$ is that for higher $m$, the growth rate is
larger over the plotted ($k_{crit}/|k|$, $Q^2$) plane than for low $m$,
and for high $k_{crit}/|k|$, the growth rate remains relatively large
for moderate values of $Q^2$ above the line $\nu_{BLLT} = 0$. This
enhanced growth at high $m$ is because the $J$-parameter is
proportional to $m$ and is contributing to the higher order terms in
the dispersion relation.

Note that there is a kink in the lower right corner ($k_{crit}/|k|
\approx $ 1.6) of the $m=$2 contour plot for the real component of the
root. This occurs because in adjacent regions to the kink different
real components have the most negative imaginary component.

Figure 2 and table 1 also indicate that the greatest growth occurs for
small values of $Q^2$, just as predicted using the BLLT dispersion
relation (cf. Sect. \ref{sect:acoustic}). Moreover, they indicate that
the disk is unstable to form spirals for a wide range of $Q$ and $m$,
although the growth rate is low, of order $\epsilon$, when $Q$ is
large. This implies there is still a spiral instability at low
gravity. For most bright galaxies, however, the region where the
rotation curve is flat is also the region where $Q$ is relatively
small, so these high $Q$ solutions are not important. They could be
important in early type galaxies (Caldwell et al. 1992) or low surface
brightness galaxies (van der Hulst et al. 1993) where $Q$ is high in
the main disk.

\subsection{Exponential self-gravitating disk with solid body rotation}
\label{sect:sgsolid}

The inner parts of galaxies and small galaxies typically have rotation
curves that are approximately solid body. This is the result of a
strong bulge with a nearly uniform central density in some spiral
galaxies, and a relatively dense dark matter halo in dwarf galaxies.
Inner galaxy disks (Elmegreen et al. 1998) and dwarfs (Hunter et al.
1998) may also be weakly self-gravitating for some time (e.g., between
accretion events and starbursts), and so the high-$Q$ cases studied
here may have applications there.  Furthermore, inner disks and dwarfs
have short rotation times, so the actual growth factor of a spiral
instability can be large even if the normalized growth rate is small.

For solid body rotation, 
$r\,{\Omega}^{'}/\Omega\,=\,r\,{\kappa}^{'}/\kappa\,=\,0$, and
$\Omega/\kappa\,=\,1/2$. We assume a value for $\epsilon = 0.1$ 
as in the previous section. In 
this case the term $A(r)$ does not depend on $\nu$
and $B(r)$ has a $1/\nu$ dependence. The dispersion relation then becomes 
cubic in $\nu$:
\begin{equation}
{\nu}^3 + \left(-1 + D_2 + a_3\,D_1\right)\,\nu + 
\,b_2\,D_0 = 0, \label{eq:ncubic} \nonumber
\end{equation}
where the terms $D_i$, $a_3$, and $b_2$ were defined in 
the previous section.
The roots can be expressed as an expansion in $\epsilon$, writing $\nu = {\nu}_0 + 
{\nu}_{1}\,\epsilon + {\nu}_{2}\,{\epsilon}^2 + ... $. The zero order term is 
the Lin-Shu dispersion relation, equation (\ref{eq:linshu}), with 
$\eta$ replaced by $\hat{\eta}$. 
The first order term is
\begin{equation}
\nu_1\,\epsilon =  -i\,\epsilon\,{{\cos{p}}\over{2\,\nu_0}}\left[{{Q^2}\over{4\,
\hat{\eta}}}\left(1 - {{r {\sigma_0}^{'}}\over{\sigma_0}} - 2\,\sin^2 
{p}\right) -{{1}\over{2}} -{{r\,{\sigma_0}^{'}}\over{\sigma_0}} + 
{{\sin^2 {p}}\over{2}}\right]. 
 \label{eq:cubicg} 
\end{equation}
In the region where $\nu_0$ is real, the growth rate is dominated by the first order term.
In the region where $|{\nu}_{0}|$ is of order
1, $\nu_1 \sim -i$ and the growth rate is of order $i \nu_1\,\epsilon \approx 
\epsilon$. For an exponential disk at two scale lengths ${{r {\sigma_0}^{'}}
\over{\sigma_0}} = -2$, so 
\begin{equation}
\nu_1\,\epsilon = -i\,\epsilon\,{{\cos{p}}\over{2\,\nu_0}}\left[{{Q^2}\over{4\,\hat{\eta}}} 
\left(3 - 2\,\sin^2 {p}\right)
 +  {{\sin^2{p}+3}\over{2}}\right].
\end{equation}

Figure 3 and table 2 show real and imaginary components of the
normalized roots of the full dispersion relation (\ref{eq:ncubic}) for
the rising rotation curve model at $r=2\,r_{d}$. Again we display only
the root that corresponds to the fastest growth. We can see from the
left-hand regions in the ($k_{crit}/|k|,Q^2$) plot, where the absolute
values of the real components are large, that the growth rates become
small for small $k_{crit}/|k|$. The opposite occurs for small values
of the real component, which are in the lower region of the plot. Where
the real component is of order 1, in the center of the plot, the growth
rate is of order $\epsilon$.

The detailed behavior of the growth rate in this case can be followed
from the approximate analytical solution written above as equation
(\ref{eq:cubicg}). For example, equation (\ref{eq:cubicg}) gives the
same growth rate as the full solution in table 2 for $\eta =
k_{crit}/|k| = 0.2$ for both $m =$ 2 and 5, because the approximate
equation is relatively accurate for low $\hat{\eta}$.  Equation
(\ref{eq:cubicg}) gives slightly different rates than table 2 for
$\eta=0.6$; at $m$ = 2 and $Q^2$ = 2, 5, and 10, equation
(\ref{eq:cubicg}) has growth rates of 0.232, 0.228 and 0.275 while
table 2 has more precise growth rates of 0.227, 0.225 and 0.273.  The
rates given by equation (\ref{eq:cubicg}) differ more significantly
from those calculated by equation (\ref{eq:ncubic}) when $\eta>0.6$.

One can observe from table 2 that the real component corresponding to
the greatest growth rate in the ($k_{crit}/|k|, Q^2$) plane is always
negative, i.e., it corresponds to the Lin-Shu and BLLT solutions
inside of corotation in the disk.

Figures 2 and 3 show that there is a similarity between the growth
rates for the flat and solid body rotation curve models. Both figures
display a saddle shape for the growth rate contours; the greatest
growth occurs as $Q^2<1$, and for $k_{crit}/|k| \approx 0.5$, the
growth rate first decreases and then increases with increasing $Q^2$.
The main difference between the two models occurs for large numbers of
spiral arms, where the growth rate is smaller at $m=5$ than $m=2$ for
the solid body case, and larger at $m=5$ than $m=2$ in the flat
rotation curve case. This is because for solid body rotation, $J= 0$,
so the absence of differential rotation reduces the growth rate of
waves at any $m \ne 0$. The zero order BLLT instability condition (Eq.
(\ref{eq:lbunstable})) is reduced to the Lin-Shu instability condition
(Eq. (\ref{eq:linshu})), and the acoustic instability disappears as the
upper unstable region collapses around $\nu^2 \approx 1$. In addition,
the contributions of $J^2$ to the higher order terms $\nu_i$ are also
absent so the growth rate is less than for $J^2 >0$.

The solutions shown for all the self-gravitating models indicate that
disks are weakly unstable to spiral waves when $\epsilon =
1/(k_{crit}r) > 0$, even in the limit of weak self-gravity.  This is
the first time spiral disk instabilities have been found at large $Q$
in the absence of magnetic fields.  We pursue this result further in
the next section, which considers the growth of waves in the absence of
self-gravity, that is, when $\epsilon = 0$.

\subsection{Exponential disk with solid body rotation and no self-gravity}
\label{sect:nsgsolid}

This section and the next consider fluid disks without self-gravity
as an idealization of the high $Q$ cases found to be unstable in the
previous sections. To be consistent with the radial dependence of the
enthalpy amplitude, $H(r)$, used before, which was defined in terms of a
slowly varying potential amplitude $\Phi$, we now assume $H(r) \propto -f(r)$, where
$f$ was given in the discussion following equation (\ref{eq:poisson}).

From equation (\ref{eq:cubicg}) we can see that the growth rate of the
instability depends on both the self-gravity of the disk and the radial
derivative of the background surface density. The normalized
growth rate is, to first order, ${\nu}_1\,\epsilon$, from the previous
discussion.  The first term in equation (\ref{eq:cubicg}) is proportional
to $Q^2 \cos{p} \epsilon/(4 \hat{\eta}) = k\, {a}^2/{\kappa}^2\,r$, which is
independent of the self-gravity of the disk. It depends primarily on
the disk curvature, i.e., on the ratio of the square of the semimajor
axis of an epicycle caused by random motions ($a/\kappa$), to the
product of the wave scale ($k^{-1}$) and the disk radius ($r$).  The
second term is proportional to $\epsilon \propto
{\rm mass}_{d}/{\rm mass}_{total}$, which comes from the self-gravity of the
disk. If the disk self-gravity is neglected, $\epsilon=0$ and the
second term is zero, but there is still growth from the first term, 
depending on orbital curvature.
When $\epsilon = 0$, the expansion has to be made in terms of the small
parameter $\zeta\equiv 1/\hat{k} r$. Then we get:  
\begin{eqnarray}
{{\nu}_{0}} &= &\pm  \sqrt{1 + {(a\, \hat{k})}^2/{\kappa}^2}, \nonumber \\
{{\nu}_{1}\,\zeta} &= & i\,\frac{k\,{a}^2}{2\, {\kappa}^2\,r\,{\nu}_{0}}
\left(2\,\sin^2 {p} - \frac{r}{r_{d}} - 1\right). \label{eq:solid1} 
\end{eqnarray}
When the expression inside the parenthesis of equation (\ref{eq:solid1}) is zero for
some particular pitch angle $\pi-p$, there is no growth at that
radius, but there is growth at adjacent radii.

The numerical solutions to equation (\ref{eq:ncubic}) when gravity is
neglected are shown for $r = 2\,r_{d}$ in figure 4 and table 3, using
normalized axes $(a/\kappa r )^2$ instead of $Q^2$ and $1/|kr|$ instead
of $k_{crit}/|k|$.  To compare the growth rates with the previous
models, recall that the value of the vertical axis in figure 4 is
obtained by multiplying the value of the vertical axis in our previous
figures by ${\epsilon}^2$ = 0.01, and the value of the horizontal axis
in figure 4 is obtained by multiplying the previous value of the
horizontal axis by $\epsilon$=0.1. This means that the growth rates in
figure 4 are analogous to those in the upper right part of figure 3.
Figure 4 and table 3 indicate that the growth rate remains finite,
proportional to $(a/\kappa r )$, as $|kr| \rightarrow \infty$. We infer
from this behavior that the instability is acoustic in nature, similar
to that described in section \ref{sect:acoustic}, 
but in the absence of self-gravity
and shear. It is driven by curvature and pressure gradients in the disk
(cf. section \ref{sect:bessel}).

\subsection{Exponential disk with Keplerian rotation and no self-gravity}
\label{sect:nsgkepler}

Accretion disks around black holes (Nakai et al. 1993) and protostars 
have negligible self-gravity and may have Keplerian rotation. In
this case $r \kappa^{'}/\kappa = r \Omega^{'}/\Omega = -3/2$, and
equation (\ref{eq:operator}) becomes a fifth order polynomial in $\nu$,
as for a flat rotation curve. As in the previous section, an acoustic
instability is still present even in the absence of self-gravity. The
dispersion relation for this acoustic instability is now obtained from
equation (\ref{eq:disp}) with the modifications
\begin{eqnarray*}
D_0 &= &\delta \\
D_1 &= &\delta\left( i k r + \frac{r f^{'}}{f}\right) \\
D_2 &= &\delta \left(-{\hat{k}}^2 r^2 + 2 i k r \frac{r {f}^{'}}{f} + 
\frac{r^2 f^{''}}{f}\right). \\
\end{eqnarray*}
Because there is no self-gravity, this fifth order dispersion relation has to 
be expanded to successive orders in 
$\zeta=1/{\hat k}r$ instead of $\epsilon$, giving $\nu=\nu_0+\nu_1\zeta+
\nu_2\zeta^2 ...$. 
The zero-order term in this expansion is the modified BLLT dispersion 
relation, equation (\ref{eq:lb2}).
Recall that the geometric term for a Keplerian disk is ${s} = \sqrt{6}$. 
The zero order term becomes complex when the instability condition, equation 
(\ref{eq:shear}), is satisfied. When equation (\ref{eq:shear}) is not satisfied,
the growth rate is dominated by the first order term
\begin{equation}
\nu_1 \zeta = -i \frac{a^2 k}{2 \kappa^2 r}  
\frac{\left(1 + r/r_{d} - 2 \sin^2 {p} \right)
\left({\nu_0}^2 - 1\right) - 3\left(1 + m \nu_0 \right)}
{ \nu_0 \left(2 {\nu_0}^2 - 2 - a^2 \hat{k}^2/\kappa^2 \right)} .
\label{eq:kepler6}
\end{equation}

Figure 5 and table 4 display numerical solutions to the fifth order
polynomial, equation (\ref{eq:disp}), for the dispersion relation in
this Keplerian model using the modified expressions $D_{i}$ when
self-gravity is neglected. The real and imaginary components of the
root with the largest growth rate are plotted using the same axes as in
the previous section, ($1/|kr|$, $a^2/\kappa^2 r^2$). The critical
curve for stability, equation (\ref{eq:kepler7}), is plotted in the top
figures as a thick line.  To the left of the critical curve, equation
(\ref{eq:kepler7}) is not satisfied and the disk is stable against
acoustic instabilities to lowest order (higher order instabilities
remain).  To the right of the critical curve, equation
(\ref{eq:kepler7}) is satisfied and the acoustic instability to all
orders dominates the growth of perturbations. The growth rate is larger
than that in the case of solid body rotation without self-gravity
because shear stimulates growth. There are discontinuities in the
contours for the real component, with kinks at the same locations in
the contours of the imaginary growth rate near the critical curve. 
To the left of these discontinuities, the real part of $\nu$ is negative,
corresponding to radii inside the ILR; to the right, the real part
is positive, corresponding to radii outside the OLR. 

Equation (\ref{eq:kepler6}),
which is the first order approximation to the growth rate, matches the
full numerical solutions in figure 5 and table 4 to two significant
digits for $\hat{k} r > 5$ and $(a^2/\kappa^2 r^2) < 0.1$.

\section{Physical Insights to the Curvature Terms}
\label{sect:bessel}

We have just shown that differential rotation, curvature, and radial
gradients in the basic properties of a fluid disk affect the
propagation and growth of spiral disturbances. 
Here we simplify the problem by including only the
effects of orbital curvature.

The curvature terms can be illustrated by
considering an ideal disk with solid-body rotation, constant 
surface density, and negligible self-gravity. Such disks
may be appropriate for the central regions of quiescent galaxies,
such as NGC 2207 (Elmegreen et al. 1998).
The governing equation (\ref{eq:operator}) for such 
a disk, assuming constant surface density, becomes
\begin{equation}
\frac{d^2h_1}{dr^2}+\frac{1}{r}\frac{dh_1}{dr}
+\left(\frac{{\kappa}^2}{a^2}\left[{\nu}^2-1\right]-\frac{m^2}{r^2}\right)h_1=0.
\label{eq:bessel}
\end{equation}
This equation was derived with the center of the coordinate system
at the center of rotation.  It is the well-known Bessel equation, and
can have mathematical singularities at $r=0$.  Often in wave
equations, these mathematical singularities can be transformed away by
a change in the coordinate system, adopting, for example, a rectilinear
coordinate system instead of cylindrical.  However, in the case of a
galaxy, the singularities cannot be transformed away by a different
coordinate system: rotation and galactic gravity define the coordinate
system. 

In the galactic Bessel equation, the time derivative in the equation of
motion appears as the term $\nu^2$, as it did in the previous
sections.  The spatial variation in the azimuthal direction is also
assumed to be the same as before, $e^{-im\theta}$, but in the radial
direction it is written explicitly.  We may look for the behavior of
{\it spirals} by assuming radial solutions of the form $h_1 \propto
e^{i k r}$.  These solutions are trailing spirals when $k<0$.  They
always contain pieces of waves that can come close to the origin, 
depending on their direction of propagation, so
they can force out the singularities in the Bessel equation.  The pure-ring
case with $m=0$ may also approach the origin and increase in amplitude. 
In this case the increase
is analogous to laboratory
sonoluminescence, in which sound waves converge to the
center of an air bubble in a liquid and increase in amplitude until they
shock and emit light (Kondic, Gersten, \& Yuan 1995).

In the case of spiral solutions, the radial derivatives in the Bessel
equation are replaced by $ik$ and the frequency $\nu$ may be solved to
give 
\begin{equation}
{\nu}^2=\left[\left(\frac{a}{{\kappa}r}\right)^2\left(r^{2}k^{2}+m^2\right)
+1\right]-irk.
\label{disp2}
\end{equation}
This frequency is necessarily complex because of the first derivative
term in equation (\ref{eq:bessel}).  Because of this term, the general
solutions are growing or decaying oscillations with spiral shapes
having $m$ arms.  It will become apparent shortly that the incoming
waves are growing, and the outgoing waves are decaying, as expected
from the nuclear singularity. 

For $\nu_R>>\nu_I$, we recover the same result as equation (\ref{eq:solid1}) 
in the limit $r_{d} \rightarrow \infty $, assuming 
constant enthalpy amplitude:
\begin{equation}
\nu_R={\pm}i\left[\left(\frac{a}{{\kappa}r}\right)^2\left(r^{2}k^{2}+m^2\right)
+1\right]^{1/2},\;\;\;
\nu_I=-\frac{1}{2}\left(\frac{a}{{\kappa}r}\right)^2\frac{rk}{\nu_R}.
\label{eq:nurnui}
\end{equation}
This gives the growth rate of a wave with azimuthal wavenumber $m$
and radial wavenumber $k$.  Note that $|\nu_R|>1$ in all cases here,
which means the waves are only inside the inner
Lindblad resonance or outside the outer Lindblad resonance.  This was
the case also for the new instability solutions discussed in section
\ref{sect:acoustic}, which re-considered the BLLT equations in this new
radial limit.

The nature of the growth implied by $\nu_I$ in equation
(\ref{eq:nurnui}) should be discussed more. Recall that the assumed time
behavior of the wave in an inertial frame is $e^{i\left(kr+\omega
t-m\theta\right)}= e^{i\left(kr+\nu \kappa t\right)}$ for $\nu=
(\omega-m\Omega)/\kappa$ and $\theta=\Omega t$ following the rotation.
Also note that we have written $\nu=\nu_R+i\nu_I$.   Thus we have a
time behavior $e^{i\nu_R\kappa t}e^{-\nu_I\kappa t}$.  When $\nu_I$ is
negative, the wave grows in time.  This occurs for trailing waves only
when $\nu_R$ is the negative one of the two solutions given above,
because $\nu_I\propto -k/\nu_R$, and $k<0$ for trailing waves.
Moreover, the negative $\nu_R$ solution is an incoming trailing wave,
because the wave-like part of the solution, $e^{i\left(kr+\nu_R \kappa
t\right)}$, has constant phase for decreasing $r$ with increasing time
when $\nu_R<0$, i.e., $r=-\left(\nu_R/k\right)\kappa t= -|\nu_R/k|\kappa
t$ when both $\nu_R$ and $k$ are less than zero.  As a result, {\it the
galactic Bessel equation has trailing spiral wave solutions that grow
in time as they propagate toward the center of the galaxy.}

These solutions are not instabilities in the usual sense, because $t$
cannot be allowed to go to infinity.  The waves reach the center in
finite time, i.e., in the time $t\sim r/a$.  In this sense, the growing
solutions are like those in the galactic swing amplifier (Goldreich \&
Lynden Bell 1965; Julian \& Toomre 1966), in which spiral waves grow in
the shearing part of a disk for a finite time ($\Delta t\sim2/A$ for
Oort constant $A$).  The instabilities in a galactic nucleus are also 
not stationary waves that grow in amplitude without any change
in shape. This is because the spiral solution is always undefined at
the nucleus and can never be considered present at all radii. The waves
are only pieces of spirals, moving inward or outward with a
growth or decay in time following the wave crest, respectively. Thus
the growth is also unlike the growth of infinite plane waves in a
sheet, as might be the case for the Kelvin-Helmholtz instability, for
example.  Spiral wave growth in galactic nuclei involves
inward propagation of finite wave trains. 

The dispersion relation (\ref{disp2}) may also be regarded as an
equation for $k$, in which $\nu$ is held as a real variable. 
Then equation
(\ref{eq:bessel}) has normal Bessel function solutions
$J_m({k_B}r)$ and $Y_m({k_B}r)$ for
\[
\frac{{\kappa}^2}{a^2}({\nu}^2-1){\equiv}{k_B}^2>0  ; \] 
$k_B$ is the radial wavenumber.
When ${k_B}r$ is large, $J_m$ and
$Y_m$ behave like sines and cosines, which may be combined as outgoing or
incoming waves with $\exp(i{\omega}t)$. When $k_{B} r$ is small,
the $Y_m$ solutions grow algebraically with decreasing $r$. The growth
arises directly from the curvature terms, namely, the first derivative
term and the $m^2/r^2$ term.  These are the same terms that led to
imaginary $\nu$ in equations (\ref{disp2}) and (\ref{eq:nurnui}). 
The $m^2/r^2$ term
actually defines a region within which the waves start to grow out of
bounds, i.e., when $r<m/{k_B}$ for  $m{\neq}0$, the 
$Y_m(k_Br)$ solution begins to increase.  In terms of the growth
discussed for the time-dependent case, this is the radius at which an
incoming wave has only one more epicycle in time before it reaches the
nucleus propagating at the sound speed.

What happens to a trailing spiral wave in a real galaxy when it enters
the $rk_B/m<1$ regime?  We expect that the amplitude will begin to
increase geometrically until nonlinear and dissipative effects come into
play. This means that the waves will break in the form of shocks
shortly after they enter the inner region.  The condition $rk_B/m<1$
implies that the radius for this wave shocking increases with azimuthal
wave number $m$.  This explains for the case of NGC 2207 (Elmegreen et
al. 1998) why the multiple-arm features are only observed in the outer
part of the nuclear disk, while the $m=1$ and ring-like feature is
close to the center.  That is, the multiple arms (high $m$) become
non-linear and damp out before they reach the inner radii, leaving only
the low-$m$ arms near the center.  Other spirals that may travel
outward in NGC 2207 are probably too weak to be seen because their
amplitudes decrease as they propagate.

Galactic nuclear spiral waves also propagate in the azimuthal
direction with angular speed ${\omega}/m$ as long as ${\nu}^2-1>0$. This 
angular speed
implies that waves with different $m$ will interact, forming complex 
structures.  The waves are also dispersive, with dispersion relation
\[
\frac{\omega-m\Omega}{\kappa}=
{\pm}\left(1+\frac{a^2{k_B}^2}{\kappa^2}\right)^{1/2}.
\]
They form wave packets that propagate with group velocity
\[
  c_g={\pm}a\sin{\gamma_B},
\]
with $\gamma_B = \tan^{-1} a k_B/\kappa$. Undoubtedly the waves will
interact because of these various phase and group speeds. They will
also get sheared by differential rotation in reality to form complex
spiral structures. When $\nu^2-1<0$, the entire disk is evanescent;
then we should not see any waves.

So far we have ignored the exponential 
density distribution of the disk. If 
it is taken into consideration, equation (\ref{eq:bessel}) will 
change to
\begin{equation}
\frac{d^2h_1}{dr^2}+\frac{1}{r}\left(1-\frac{r}{r_{d}}\right)\frac{dh_1}{dr}
+\left[\frac{{\kappa}^2}{a^2}\left({\nu}^2-1\right)-\frac{m^2}{r^2}+\frac{m}
{{\nu}r r_{d}}\right]h_1=0,
\label{exp}
\end{equation}
where $r_d$ is the scale length of the exponential disk. The additional
factor in the first derivative term will modify the behavior of the
Bessel functions when $r{\ge}r_d$, and the additional term in the last
parenthesis will complicate the wave behavior. But the qualitative
nature of the Bessel function solutions does not change.

\section{Summary}

We have obtained dispersion relations for spiral waves with multiple
arms, considering curvature and gradient terms that were ignored in
previous derivations.  These dispersion relations suggest the presence
of several new instabilities.  Four specific cases
were studied, flat and rising rotation curves with self-gravity,
rising rotation curves without self-gravity, and Kepler
rotation curves without self-gravity. These cases seem to have
applications in various regions of galaxies and accretion disks.

When self-gravity is present, instability at lowest order in the
parameter $\epsilon$ (cf. Eq. \ref{eq:epsilon}) is driven by both shear
and self-gravity. Then there are two independent instability
conditions, either of which can cause spiral waves. These are equations
(\ref{eq:lbunstable}) and (\ref{eq:shearg}).  The first of these comes
from Bertin et al. (1989), and contains the Toomre (1964) instability
condition, $Q<1$, as a special case for ring-like perturbations ($m=0$,
which gives $J=0$).  This first instability is the spiral instability
that is commonly discussed in the literature as a source of multiple
arm and grand design spiral structure in galaxy and protoplanetary
disks.  The second of these conditions arises outside the Lindblad
resonances from a combination of parameters different than the first
when $Q^2>4{\hat \eta}$ (cf. Eq.  \ref{eq:lbshear}).  When self-gravity
is not present, this second case is still unstable as a result of
pressure and differential rotation alone, as determined by the
smallness of the parameter $a/(\kappa r)$ (cf. Eq.  \ref{eq:kepler7}).
This pressure-rotation instability is apparently new, and we call it an
{\it acoustic} instability.
A physical explanation for it was given in section \ref{sect:blltextend}.

We also found additional instabilities coming from higher order terms
in an expansion of the dispersion relation (\ref{eq:disp})
around the small parameter $\epsilon$. These additional instabilities
are present even when the BLLT and Toomre instability conditions
are not satisfied, i.e., when the low order terms give stability.  The
source of these residual instabilities is a combination of orbital
curvature [terms of order $1/(kr)$], self-gravity (terms of order
$\epsilon$), and various disk gradients ($r\sigma^\prime/\sigma$,
$ra^\prime/a$, etc.), including shear (the $J$ or ${s}$ terms).
Growth rates for these residual instabilities were given to all orders
in $\epsilon$ for flat and rising rotation curves by figures 2 and 3
and tables 1 and 2, and they were given to first order in $\epsilon$ by
equation (\ref{eq:cubicg}) for solid body rotation.  The residual
instability that arises from self-gravity and orbital curvature
(through $\epsilon$), discussed in sections \ref{sect:sgconstant} 
and \ref{sect:sgsolid}, will be called a
{\it gravitational-curvature} instability. The residual
instability that arises from a combination of pressure and orbital
curvature [through $a/(\kappa r)$], discussed in sections \ref{sect:nsgsolid}
and \ref{sect:nsgkepler},
will be called an {\it acoustic-curvature} instability,
because it operates even without self-gravity. 

These three new instabilities should be important 
for fluid disks with negligible or weak self-gravity, including
proto-planetary disks, gaseous disks around
black holes, some galactic nuclear disks, low surface
brightness galaxy disks, and some dwarf galaxies. 
In these cases,  zero-order acoustic and higher-order acoustic-curvature 
and gravitational-curvature instabilities can lead
to the growth of spiral or other structures in about an orbital time. 
They are most important in the region close to the center where
the orbital time is small.

Non-linear effects arising from these waves 
may ultimately lead to visible dust lanes (Elmegreen et al. 1998)
and associated
gaseous shocks (Roberts 1969) in even the most 
weakly self-gravitating disks, with the possibility of heightened
self-gravity and star formation in some of the compressed regions
(e.g., Elmegreen 1994).
Non-linear effects might also promote
accretion flows (e.g., Larson 1990).
Indeed, the ubiquity of acoustic waves in disks implies
that galactic nuclear accretion should occur in a wide variety
of environments with or without 
shear, self-gravity, or magnetic fields. 



\newpage

\begin{deluxetable}{lccccccccccc}
\scriptsize
\tablecaption{Flat rotation curve at $r = 2\,r_{d}$}
\tablewidth{0pt}
\startdata
$Q^2$$\;\;\;$$\frac{k_{crit}}{|k|}=$ &0.2 &0.6 &1.0 &1.4 &1.8 & &0.2 &0.6 &1.0 &1.4 &1.8 \nl

&&$m = 2$&growth rate&&&&&$m = 2$&frequency&& \nl

10 &0.280 &0.332 &0.340 &0.325 &0.302&&-7.660&-2.547&-1.644&-1.318&-1.164 \nl
 5 &0.218 &0.272 &0.264 &0.236 &0.214&&-5.229&-1.709&-1.156&-0.973&-0.892 \nl
 2 &0.175 &0.236 &0.213 &0.184 &0.168&&-2.924&-0.864&-0.673&-0.658&-0.664 \nl
 1 &0.188 &0.383 &0.221 &0.163 &0.147&&-1.513&-0.297&-0.402&-0.510&-0.568 \nl
0.1&1.867 &0.875 &0.410 &0.179 &0.130&&-0.042&-0.054&-0.116&-0.301 &0.457 \nl

&&$m = 5$&growth rate&&&&&$m = 5$&frequency&& \nl

10 &0.319 &0.421 &0.453 &0.446 &0.427&&-7.695&-2.653&-1.797&-1.493&-1.348 \nl
5  &0.258 &0.364 &0.358 &0.309 &0.262&&-5.255&-1.787&-1.256&-1.058&-0.942 \nl
2  &0.215 &0.275 &0.322 &0.358 &0.385&&-2.941&-0.853&-0.529&-0.420&-0.372 \nl
1  &0.224 &0.581 &0.559 &0.544 &0.540&&-1.531&-0.238&-0.229&-0.241&-0.249 \nl
0.1&1.914 &1.057 &0.810 &0.714 &0.673&&-0.044&-0.073&-0.116&-0.156&-0.183 \cr

\enddata

\end{deluxetable}
\newpage

\begin{deluxetable}{lccccccccccc}
\scriptsize
\tablecaption{Solid body rotation at $r = 2\,r_{d}$}
\tablewidth{0pt}
\startdata
$Q^2$, $\;\;\;\;$$\frac{k_{crit}}{|k|}=$ &0.2 &0.6 &1.0 &1.4 &1.8 & &0.2 &0.6 &1.0 &1.4 &1.8 \nl

&&$m = 2$&growth rate&&&&&$m = 2$&frequency&& \nl

10 &0.254 &0.273 &0.268 &0.250 &0.229&&-7.658&-2.540&-1.634&-1.311&-1.163 \nl
 5 &0.193 &0.225 &0.223 &0.206 &0.188&&-5.228&-1.703&-1.155&-0.992&-0.930 \nl
 2 &0.154 &0.227 &0.211 &0.183 &0.164&&-2.922&-0.881&-0.723&-0.727&-0.748 \nl
 1 &0.173 &0.376 &0.244 &0.184 &0.158&&-1.512&-0.389&-0.508&-0.610&-0.673 \nl
0.1&1.838 &0.789 &0.361 &0.206 &0.158&&-0.052&-0.121&-0.306&-0.483&-0.594 \nl

&&$m = 5$&growth rate&&&&&$m = 5$&frequency&& \nl

10 &0.252 &0.251 &0.222 &0.187 &0.157&&-7.692&-2.636&-1.770&-1.464&-1.323 \nl
 5 &0.192 &0.208 &0.190 &0.166 &0.144&&-5.252&-1.766&-1.227&-1.056&-0.983 \nl
 2 &0.152 &0.215 &0.207 &0.186 &0.169&&-2.938&-0.901&-0.709&-0.680&-0.674 \nl
 1 &0.171 &0.406 &0.311 &0.257 &0.226&&-1.523&-0.382&-0.451&-0.507&-0.537 \nl
0.1&1.843 &0.829 &0.487 &0.372 &0.320&&-0.053&-0.135&-0.292&-0.392&-0.443 \cr

\enddata

\end{deluxetable}
\newpage

\begin{deluxetable}{lccccccccccc}
\scriptsize
\tablecaption{Solid body rotation with no gravity at $r = 2\,r_{d}$}
\tablewidth{0pt}
\startdata
$(\frac{a}{\kappa r})^2$, $\;\;\;\;$$\frac{1}{|k\,r|}=$ &0.2 &0.6 &1.0 &1.4 &1.8 & &0.2 &0.6 &1.0 &1.4 &1.8 \nl

&&$m = 2$&growth rate&&&&&$m = 2$&frequency&& \nl

1.0 &1.205 &0.482 &0.376 &0.433 &0.462&&-5.584&-2.816& 0.777& 0.917& 0.995 \nl
0.5 &0.833 &0.311 &0.308 &0.355 &0.377&&-4.024&-2.160& 0.671& 0.769& 0.820 \nl
0.2 &0.499 &0.168 &0.139 &0.137 &0.131&&-2.673&-1.608& 0.493& 0.559& 0.598 \nl
0.1 &0.328 &0.101 &0.051 &0.032 &0.022&&-2.026&-1.355&-1.294&-1.279&-1.274 \nl
0.01&0.058 &0.014 &0.007 &0.004 &0.003&&-1.148&-1.045&-1.037&-1.035&-1.034 \nl

&&$m = 5$&growth rate&&&&&$m = 5$&frequency&& \nl

1.0 &0.684 &0.179 &0.097 &0.067 &0.051&&-7.173&-5.420&-5.271&-5.230&-5.214 \nl
0.5 &0.474 &0.122 &0.066 &0.045 &0.035&&-5.147&-3.941&-3.840&-3.812&-3.801 \nl
0.2 &0.287 &0.071 &0.039 &0.026 &0.020&&-3.375&-2.657&-2.597&-2.581&-2.575 \nl
0.1 &0.192 &0.046 &0.025 &0.017 &0.013&&-2.509&-2.036&-1.997&-1.987&-1.983 \nl
0.01&0.039 &0.008 &0.004 &0.003 &0.002&&-1.252&-1.162&-1.156&-1.154&-1.153 \cr

\enddata

\end{deluxetable}
\newpage

\begin{deluxetable}{lccccccccccc}
\scriptsize
\tablecaption{Keplerian rotation with no gravity at $r = 2\,r_{d}$}
\tablewidth{0pt}
\startdata
$(\frac{a}{\kappa r})^2$, $\;\;\;\;$$\frac{1}{|k\,r|}=$ &0.2 &0.6 &1.0 &1.4 &1.8 & &0.2 &0.6 &1.0 &1.4 &1.8 \nl

&&$m = 2$&growth rate&&&&&$m = 2$&frequency&& \nl

1.0 &1.606 &1.088 &1.266 &1.365 &1.414&&-5.736&-2.960 &2.170 &2.081 &2.040 \nl
0.5 &1.220 &0.914 &1.076 &1.139 &1.171&&-4.161& 2.010 &1.818 &1.748 &1.715 \nl
0.2 &0.861 &0.739 &0.821 &0.856 &0.875&&-2.787& 1.621 &1.489 &1.439 &1.415 \nl
0.1 &0.665 &0.597 &0.645 &0.667 &0.679&&-2.121& 1.419 &1.320 &1.281 &1.262 \nl
0.01&0.249 &0.236 &0.242 &0.246 &0.248&&-1.197& 1.102 &1.068 &1.055 &1.048 \nl

&&$m = 5$&growth rate&&&&&$m = 5$&frequency&& \nl

1.0 &1.413 &0.961 &1.097 &1.140 &1.160&&-7.262 &4.327 &4.007 &3.885 &3.821 \nl
0.5 &1.214 &1.266 &1.355 &1.382 &1.395&&-5.220 &3.253 &3.076 &3.008 &2.972 \nl
0.2 &1.041 &1.231 &1.276 &1.291 &1.298&&-3.450 &2.394 &2.291 &2.250 &2.228 \nl
0.1 &0.932 &1.089 &1.117 &1.126 &1.130&&-2.607 &1.960 &1.888 &1.859 &1.843 \nl
0.01&0.506 &0.539 &0.544 &0.546 &0.547&& 1.370 &1.229 &1.205 &1.195 &1.190 \cr

\enddata

\end{deluxetable}

\clearpage


\begin{figure}
\begin{caption} {Regions of instability in a self-gravitating disk with
constant rotation for 5 arm spirals ($m=5$). The lowest (thin) line borders
the instability condition obtained from the Lin-Shu dispersion relation
(Eq. (\protect\ref{eq:linshu})) for $m$ and $J^2 =0$. The three upper lines
bracket regions of zero-order instabilities obtained in the Lau-Bertin
dispersion relation. The border of the Lin-Shu region of instability shifts to
the lower thick line 
for nonzero $J^2$. This line is from Eq.
(\protect\ref{eq:lbunstable}) for the usual Lau-Bertin stability condition; the
two upper thick lines are the boundaries enclosing the acoustic instability
according to Eq. (\protect\ref{eq:lbshear}). The shading of the unstable
regions gives an indication of the growth rate. The most unstable
region is in the bottom left corner of the figure.  }
\end{caption} 
\end{figure} 

\begin{figure} 
\begin{caption} {Contours
showing the maximum growth rates (top) and corresponding frequencies
(bottom) for instabilities to all orders of the small parameter
$\epsilon$ in a self-gravitating disk with a constant rotation velocity
and an exponential density profile, evaluated at two scale lengths.
Solutions for two arm spirals ($m=2$) are on the left, and for 5 arm
spirals ($m=5$) are on the right.  The thick lines are obtained as in Fig. 1.,
they
border the regions of zero-order instability for the Lau-Bertin dispersion
relation. The bottom thick line is from Eq. (\protect\ref{eq:lbunstable}) for the
usual Lau-Bertin stability condition; the middle and upper lines are
the boundaries enclosing the acoustic instability according to Eq.
(\protect\ref{eq:lbshear}).} 
\end{caption} 
\end{figure} 

\begin{figure}
\begin{caption} {Same as Fig. 2 for a self-gravitating, exponential
disk at two scale lengths, but now with solid body rotation.  The thick
lines on the top figures border the regions of instability for the
zero-order Lau-Bertin dispersion relation in the case with no shear
($J=0$ in Eq. (\protect\ref{eq:lbunstable})).} 
\end{caption} 
\end{figure}

\begin{figure} 
\begin{caption} {Same as Fig. 3, but to all orders of
the small parameter $1/|k r|$ in the absence of self-gravity.  All of
the unstable growth exhibited in these solutions is from high order
terms. } 
\end{caption} 
\end{figure}

\begin{figure} 
\begin{caption}
{Growth rates and frequencies for a non-self-gravitating disk, as in
Fig. 4, but with Keplerian rotation.  The thick lines on the top
figures border the regions of instability for the zero-order,
Lau-Bertin dispersion relation in a Keplerian disk without self-gravity
(Eq. \protect\ref{eq:kepler7}).  } 
\end{caption} 
\end{figure} 
\clearpage

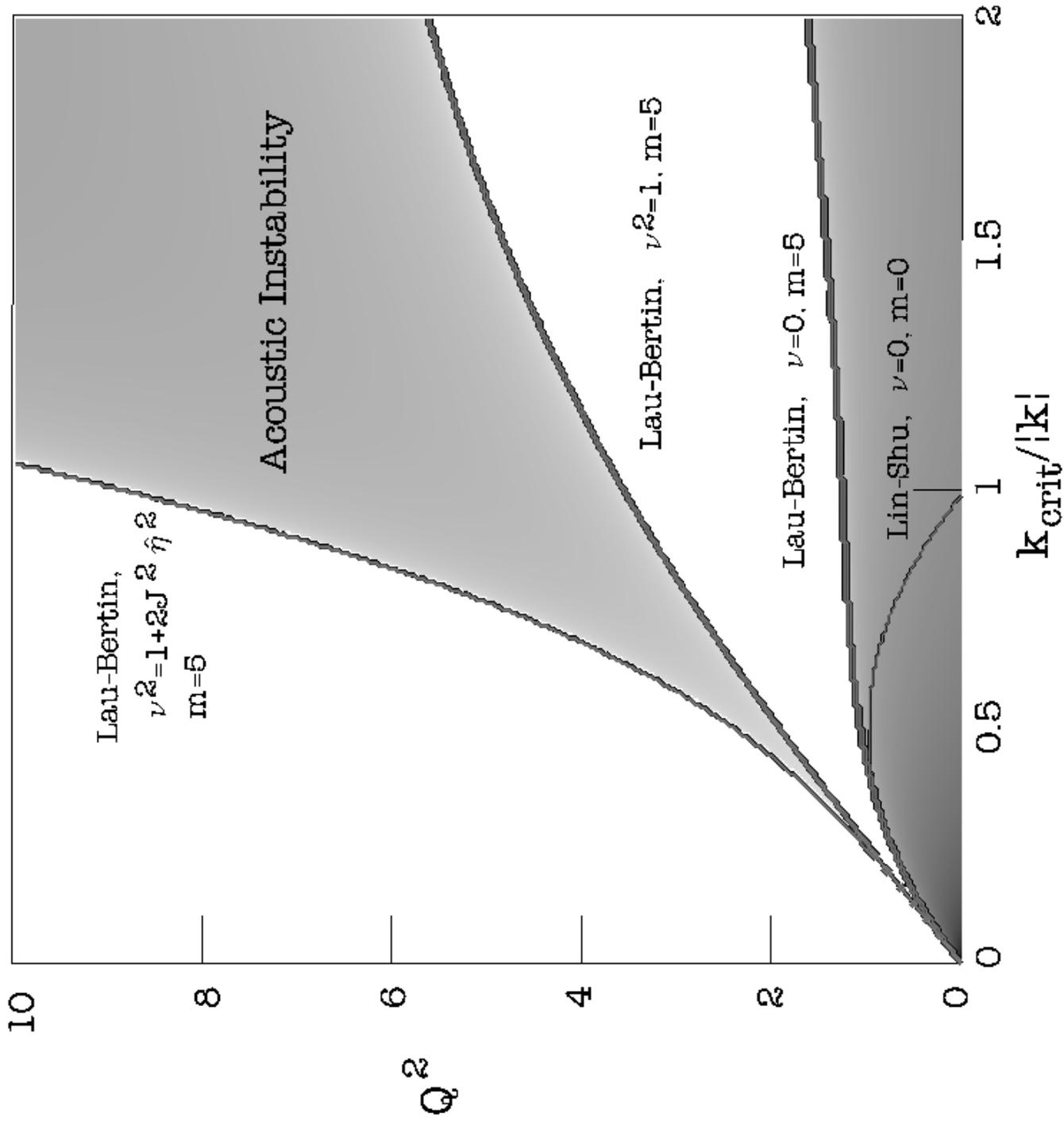
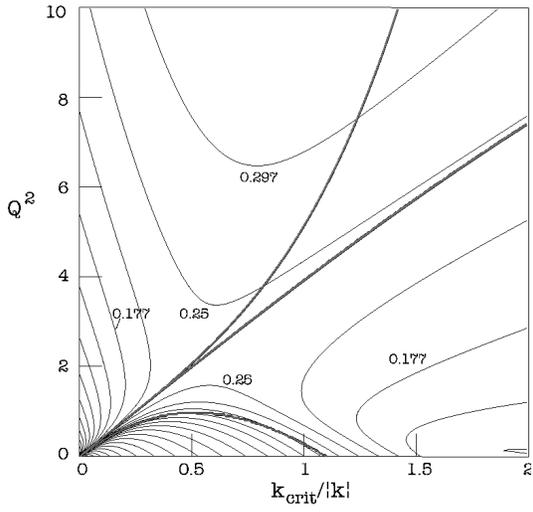
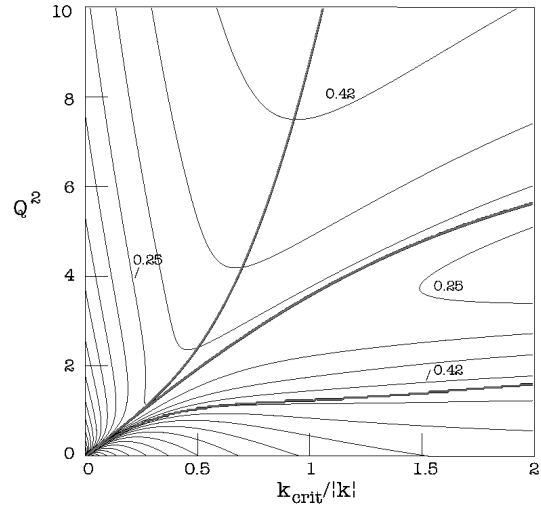
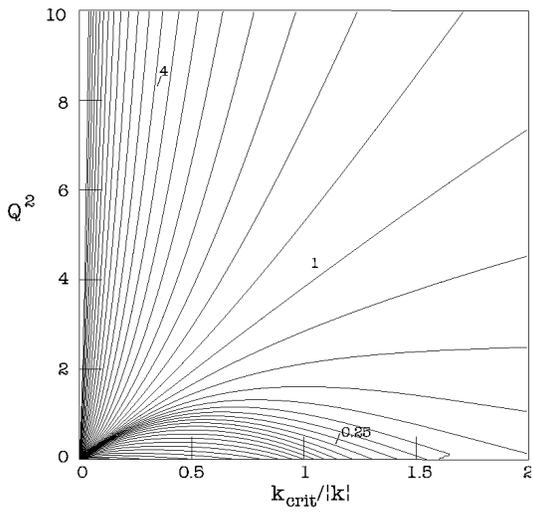
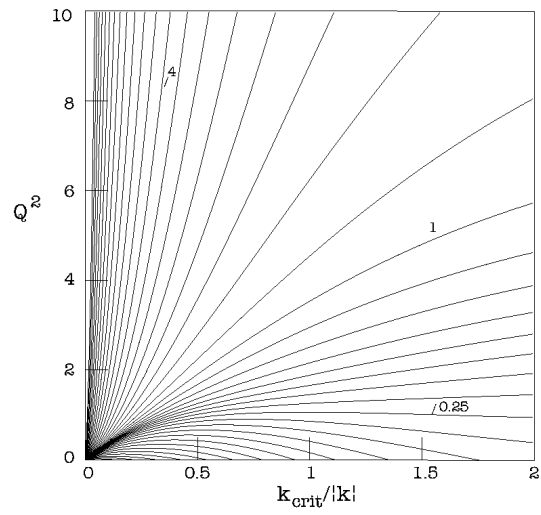
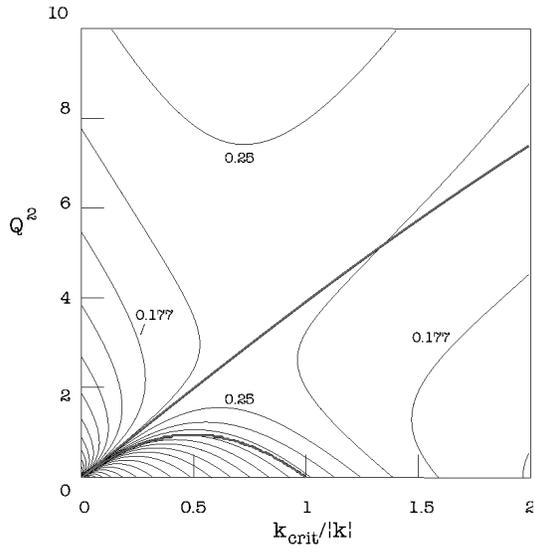
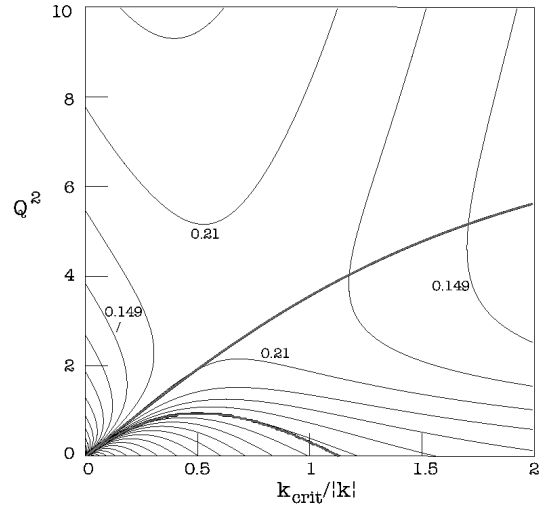
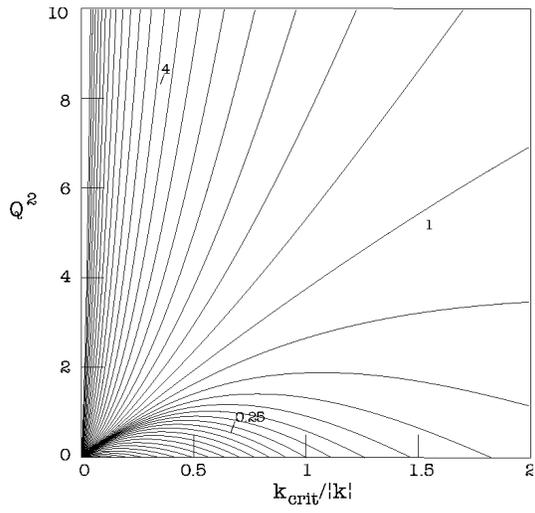
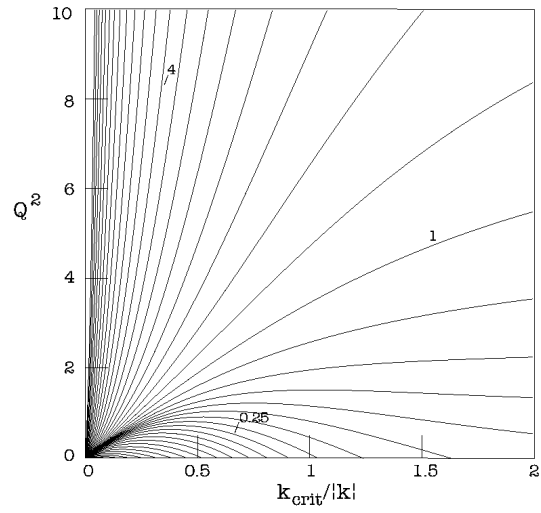
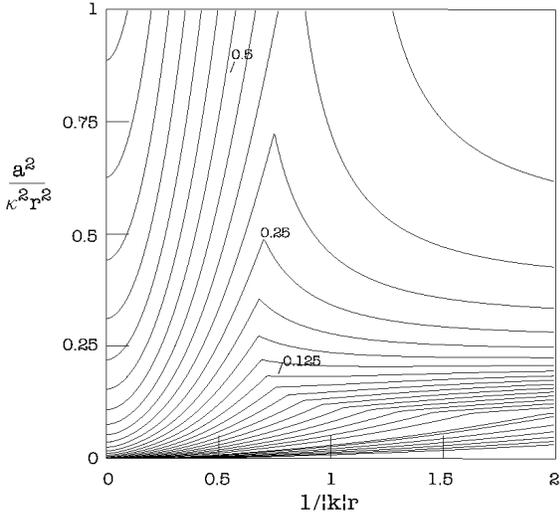
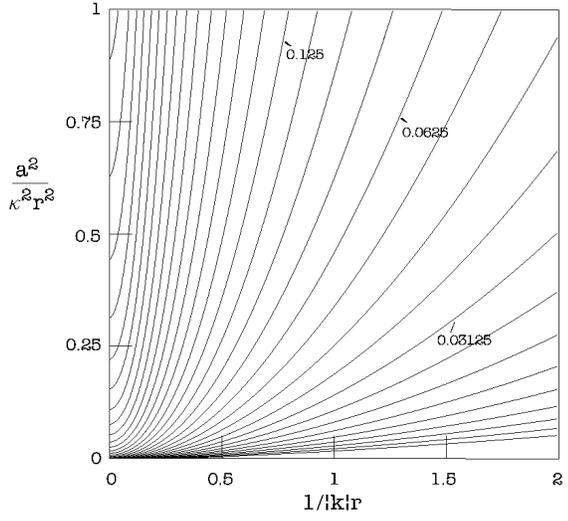
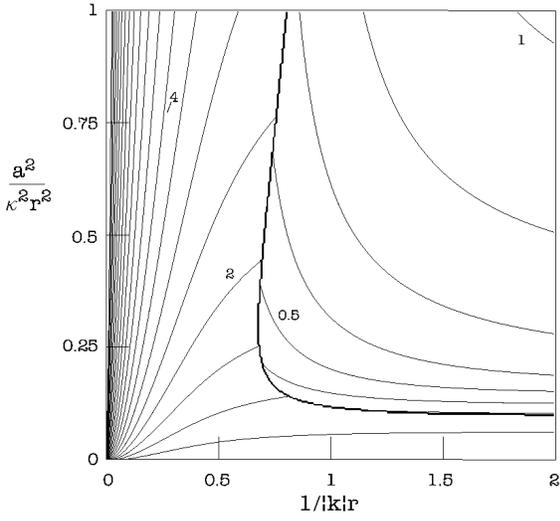
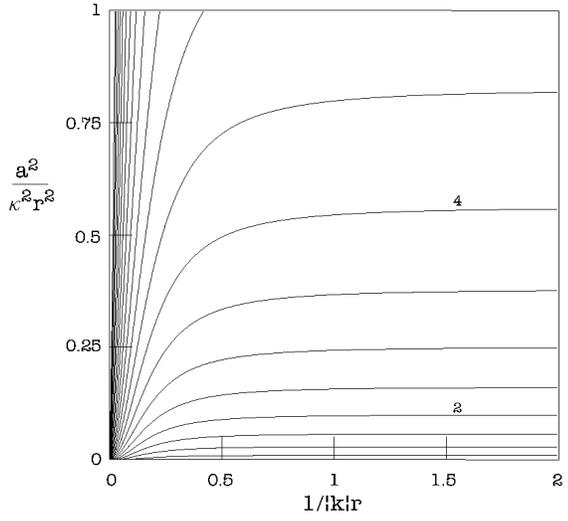
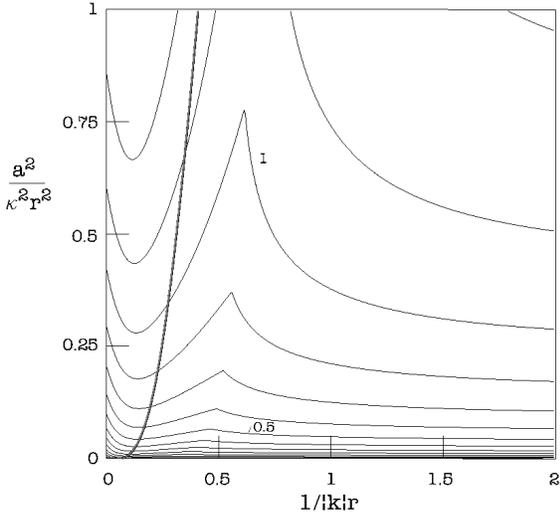
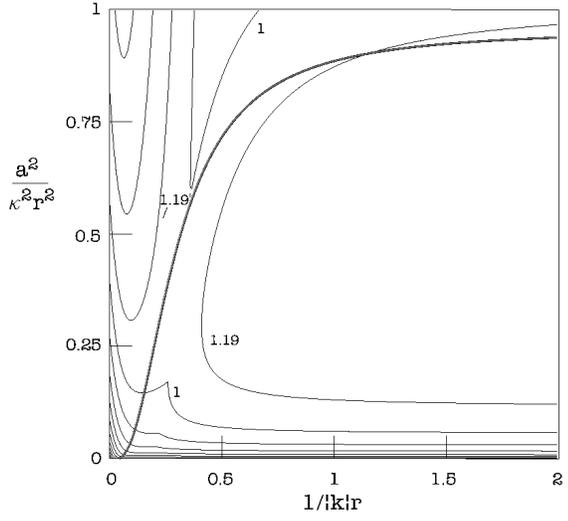
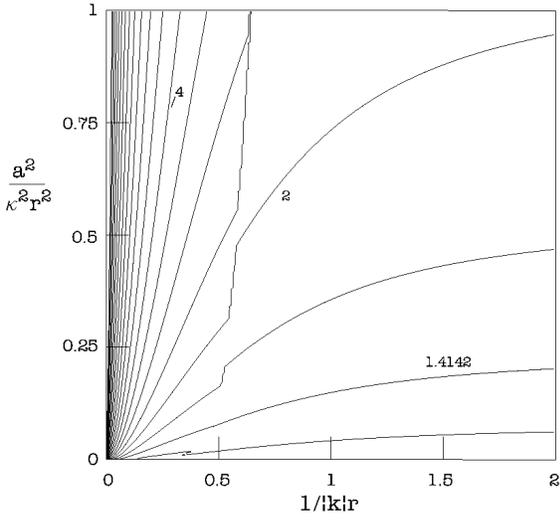
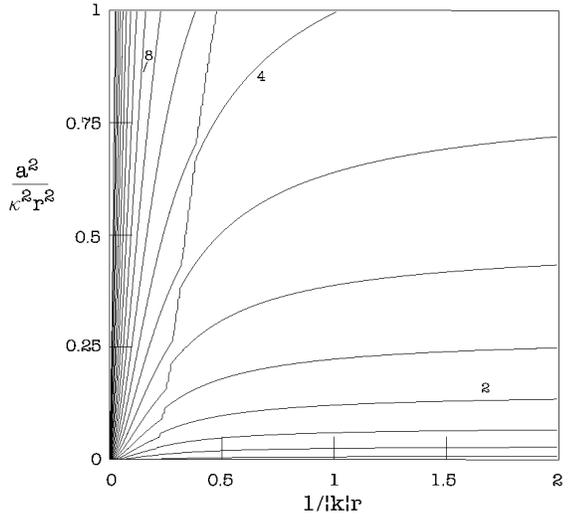
\end{document}